\documentclass[twocolumn,preprintnumbers,amsmath,amssymb,superscriptaddress,nofootinbib,longbibliography, prb]{revtex4-2}

\usepackage[utf8]{inputenc}
\usepackage[english]{babel}

\usepackage{graphicx}

\usepackage{amsfonts,amssymb,amsmath}
\usepackage{mathtools}

\usepackage{bm}
\usepackage{times}
\usepackage{bbm}
\usepackage{units}
\usepackage{dsfont}
\usepackage{mathrsfs}

\usepackage{array}
\usepackage{tabularx}
\usepackage{multirow}

\usepackage{hyperref}
\hypersetup{colorlinks=true,linktoc=all,linkcolor=blue,breaklinks=true,citecolor=blue,urlcolor=blue}

\usepackage[normalem]{ulem}
\usepackage{xspace}
\usepackage[dvipsnames]{xcolor}

\addto\captionsenglish{}

\AtBeginDocument{%
    \newwrite\bibnotes
    \def\bibnotesext{Notes.bib}
    \immediate\openout\bibnotes=\jobname\bibnotesext
    \immediate\write\bibnotes{@CONTROL{REVTEX42Control}}
    \immediate\write\bibnotes{@CONTROL{%
            apsrev42Control,author="08",editor="1",pages="1",title="0",year="1"}}
    \if@filesw
    \immediate\write\@auxout{\string\citation{apsrev42Control}}%
    \fi
}%

\newcommand*{\e}{\text{e}}
\newcommand*{\kB}{k_\text{B}}
\renewcommand*{\P}{\mathbf{P}}
\newcommand*{\bP}{\mathbf{\bar{P}}}
\newcommand*{\C}{\text{C}}
\newcommand*{\D}{\text{D}}
\renewcommand*{\H}{\text{H}}
\newcommand*{\M}{\text{W}}
\renewcommand*{\L}{\text{L}}
\newcommand*{\R}{\text{R}}
\newcommand*{\Jin}{J^Q_\text{in}}
\newcommand*{\JC}{J^Q_\C}
\newcommand*{\JH}{J^Q_\H}
\newcommand*{\JL}{J^Q_\L}
\newcommand*{\JR}{J^Q_\R}
\newcommand*{\Pcool}{\mathcal{P}_\text{cool}}
\newcommand*{\Scool}{\mathcal{S}_\text{cool}}
\newcommand*{\cI}{I}
\newcommand*{\cC}{\mathcal{C}}
\newcommand*{\bcC}{\tilde{\mathcal{C}}}
\newcommand*{\CC}{\cC_\C}
\newcommand*{\bCC}{\bcC_\C}
\newcommand*{\CH}{\cC_\H}
\newcommand*{\bCH}{\bcC_\H}
\newcommand*{\CHC}{\cC_{\H\C}}
\newcommand*{\bCHC}{\bcC_{\H\C}}

\newcommand*{\FC}{F^\C}
\newcommand*{\FH}{F^\H}

\newcommand*{\GC}{\Gamma_\C}
\newcommand*{\GH}{\Gamma_\H}
\newcommand*{\GL}{\Gamma_\L}
\newcommand*{\GR}{\Gamma_\R}
\newcommand*{\UC}{U_\C}
\newcommand*{\UH}{U_\H}
\newcommand*{\bC}{\beta_\C}
\newcommand*{\bH}{\beta_\H}
\newcommand*{\bL}{\beta_\L}
\newcommand*{\bR}{\beta_\R}
\newcommand*{\eC}{\epsilon_\C}
\newcommand*{\eH}{\epsilon_\H}
\newcommand*{\eM}{\epsilon_\M}
\newcommand*{\GCop}{\GC^{0+}}
\newcommand*{\GCip}{\GC^{1+}}
\newcommand*{\GCom}{\GC^{0-}}
\newcommand*{\GCim}{\GC^{1-}}
\newcommand*{\GHop}{\GH^{0+}}
\newcommand*{\GHip}{\GH^{1+}}
\newcommand*{\GHom}{\GH^{0-}}
\newcommand*{\GHim}{\GH^{1-}}
\newcommand*{\GRoip}{\GR^{01+}}
\newcommand*{\GRiop}{\GR^{10+}}
\newcommand*{\GRoim}{\GR^{01-}}
\newcommand*{\GRiom}{\GR^{10-}}
\newcommand*{\GLoop}{\GL^{00+}}
\newcommand*{\GLoom}{\GL^{00-}}
\newcommand*{\g}{\gamma^5}
\newcommand*{\gC}{\gamma_\C^2}
\newcommand*{\gH}{\gamma_\H^2}
\newcommand*{\gCC}{\gamma_{\CC}^4}
\newcommand*{\gbCC}{\gamma_{\bCC}^4}
\newcommand*{\gCH}{\gamma_{\CH}^4}
\newcommand*{\gbCH}{\gamma_{\bCH}^4}
\newcommand*{\gCHC}{\gamma_{\CHC}^6}
\newcommand*{\gbCHC}{\gamma_{\bCHC}^6}
\newcommand*{\rC}{r_{\cC}}
\newcommand*{\rCC}{r_{\CC}}
\newcommand*{\rbCC}{r_{\bCC}}
\newcommand*{\rCH}{r_{\CH}}
\newcommand*{\rbCH}{r_{\bCH}}
\newcommand*{\rCHC}{r_{\CHC}}
\newcommand*{\rbCHC}{r_{\bCHC}}
\newcommand*{\p}{\bar{p}}
\newcommand*{\T}{\bar{T}}
\newcommand*{\dT}{\delta T}
\newcommand*{\TL}{T_\mathrm{L}}
\newcommand*{\TR}{T_\mathrm{R}}
\newcommand*{\THot}{T_\mathrm{H}}
\newcommand*{\TCold}{T_\mathrm{C}}
\newcommand*{\etaG}{\eta_\text{global}}
\newcommand*{\etaI}{\eta_\text{info}}

\newcommand*{\diff}[2]{\frac{\mathrm{d} #1}{\mathrm{d} #2}}
\newcommand*{\difc}[2]{\mathrm{d}_{#2} #1}
\newcommand*{\difcp}[2]{\partial_{#2} #1}

\DeclarePairedDelimiter{\mean}{\langle}{\rangle}
\DeclarePairedDelimiter{\cum}{\langle\!\langle}{\rangle\!\rangle}

\newcommand{\subfigref}[2]{\ref{#1}\hyperref[#1]{(#2)}}

\begin{document}

\title{Autonomous demon exploiting heat and information at the trajectory level}

\author{Juliette Monsel}
\affiliation{Department of Microtechnology and Nanoscience (MC2), Chalmers University of Technology, S-412 96 G\"oteborg, Sweden}
\author{Matteo Acciai}
\affiliation{Department of Microtechnology and Nanoscience (MC2), Chalmers University of Technology, S-412 96 G\"oteborg, Sweden}
\affiliation{Scuola Internazionale Superiore di Studi Avanzati, Via Bonomea 256, 34136, Trieste, Italy}
\author{Rafael S\'anchez}
\affiliation{Departamento de F\'isica Te\'orica de la Materia Condensada, Universidad Aut\'onoma de Madrid, 28049 Madrid, Spain\looseness=-1}
\affiliation{Condensed Matter Physics Center (IFIMAC), Universidad Aut\'onoma de Madrid, 28049 Madrid, Spain\looseness=-1}
\affiliation{Instituto Nicol\'as Cabrera, Universidad Aut\'onoma de Madrid, 28049 Madrid, Spain\looseness=-1}
\author{Janine Splettstoesser}
\affiliation{Department of Microtechnology and Nanoscience (MC2), Chalmers University of Technology, S-412 96 G\"oteborg, Sweden}

\date{\today}

\begin{abstract}
We propose an electronic bipartite system consisting of a working substance, in which a refrigeration process is implemented, and of a nonthermal resource region, containing a combination of different thermal baths.
In the working substance, heat is extracted from the coldest of two electronic reservoirs (refrigeration) via heat- and particle transport through a quantum dot.
This quantum dot of the working substance is capacitively coupled to the resource region. In such a setup, a finite cooling power can be obtained in the working substance, while the energy exchange with the resource region exactly cancels out on average. At the same time, \textit{information} is always exchanged, even on average, due to the capacitive coupling between the two parts of the bipartite system. The proposed system therefore implements an autonomous demon with fully vanishing heat extraction from the resource. Unlike macroscopic machines, nanoscale machines exhibit large fluctuations in performance, so precision becomes an important performance quantifier.
We give a comprehensive description of the thermodynamic performance of the proposed autonomous demon in terms of stochastic trajectories and of full counting statistics and demonstrate that the precision of the cooling power strongly depends on the operation principle of the device. More specifically, the interplay of information flow and counter-balancing heat flows dramatically impacts the trade-off between cooling power, efficiency, and precision. We expect this insight to be of relevance for guiding the design of energy-conversion processes exploiting nonthermal resources.
\end{abstract}

\maketitle

\section{Introduction}

Thermoelectric energy conversion at the nanoscale~\cite{Benenti2017Jun,Whitney2019Apr} opens up for new opportunities, one of the most striking ones being the possibility to implement engines exploiting information as a resource~\cite{Maruyama2009Jan,Parrondo2015Feb}.
At the same time, nanoscale engines face new challenges, one of them being that fluctuations~\cite{Blanter2000Sep,nazarov_quantum_2003,Esposito2009Dec} can be of the same order of magnitude as the actual desired outcome. Precision has therefore recently advanced to be one of the important performance quantifiers of nanoscale thermal machines~\cite{Barato2015Apr,Guarnieri2019Oct,VanVu2022Apr,Tesser2024May}.

\begin{figure}[hb!]
    \includegraphics[width=\linewidth]{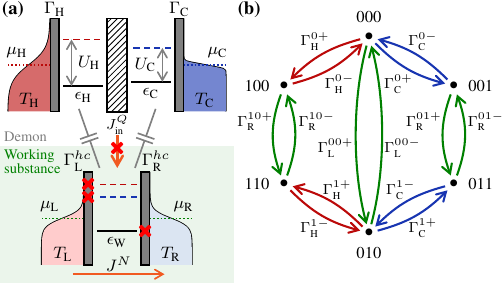}
    \vskip-2mm
    \caption{\label{fig:setup}
        (a) Setup of a single-level quantum dot, W,
        in contact with electronic reservoirs at
        temperatures $\TL$ and $\TR$ (working substance) operating as a refrigerator. The cooling power $\Pcool=J^Q_\mathrm{R}$ in the working substance is produced by exploiting a demonic resource in the absence of a steady heat flow, $J_\mathrm{in}^Q=0$. This resource consists of two quantum dots, H and C, in contact with a hot and a cold reservoir at temperatures $\THot$ and $\TCold$, capacitively coupled to each other and to quantum dot W. The remaining indicated parameters are the level energies $\epsilon_\mathrm{H},\epsilon_\mathrm{C},\epsilon_\mathrm{W}$ of the three quantum dots, the interaction energies $U_\mathrm{H},U_\mathrm{C}$ due to the capacitive coupling of the demon dots to dot W, the reservoir electrochemical potentials, $\mu_\mathrm{H},\mu_\mathrm{C},\mu_\mathrm{L},\mu_\mathrm{R}$, and the tunnel coupling strengths between dots, $\Gamma_\mathrm{H},\Gamma_\mathrm{C}$ and
        $\Gamma_\mathrm{L}^{hc},\Gamma_\mathrm{R}^{hc}$. Red crosses indicate the forbidden transitions. (b) Markov chain representing the stochastic dynamics of the triple-dot system. The states are labeled with the occupations of the dots: $hwc\in\left\{000,001,010,001,100,110\right\}$. The arrow color emphasizes which dot is involved in the particle exchange: 
        red/blue/green for H/C/W.
        The transition rates are defined in Sec.~\ref{sec:master}.
    }
\end{figure}

Recently, different types of quantum-dot setups have been proposed and experimentally implemented that operate as so-called Maxwell demons, either actively implementing feedback protocols~\cite{schaller:2011,Averin2011Dec,esposito:2012,Koski2014Sep,Schaller2018May,Annby-Andersson2020Apr,Annby-Andersson2024May,Chida2017May, Cottet2017} or operating autonomously~\cite{Strasberg2013Jan,strasberg:2018,Erdman2018Jul,Koski2015Dec}.
A Maxwell demon is a device which seemingly violates the second law, since an engine is brought to work while no energy is absorbed from the resource. In contrast, the engine exploits information and feedback as a resource, thereby re-installing the second law as required~\cite{Bennett2003Sep,Sagawa2009Jun,Parrondo2024Jul}.
Devices with capacitively coupled quantum dots~\cite{chan:2002,mcclure:2007,hubel:2007,hubel:2008,bischoff:2015,keller:2016} or single-electron transistors \cite{Averin2011Dec, Koski2014Sep, Koski2015Dec} are particularly advantageous electronic realizations of \textit{autonomous} Maxwell demons, since a capacitive coupling between them allows for a well-defined energy exchange between device elements, while at the same time a meaningful interpretation in terms of a bipartite system with resource region and working substance is guaranteed.
Other platforms for controlled coupling to quantized energy sources have been explored recently, including qubits~\cite{lloyd_quantum_1997,quan_maxwells_2006}, vibrating wires~\cite{Entin-Wohlman2010Sep,Mukherjee2020Dec,Tabanera-Bravo2024Mar}, cavity QED setups~\cite{Bergenfeldt2014Feb,Hofer2016Jan} or quantum dot refrigerators~\cite{Venturelli2013Jun}.
The mentioned proposals and realizations of autonomous Maxwell demons, however, have the drawback that the heat exchanged between working substance and demonic resource only vanishes in limiting cases, where the desired output typically also vanishes.
A different class of demonic systems has been suggested recently, so-called N-demons, which do not at all exploit information and feedback, but rather a resource that is nonthermal~\cite{Whitney2016Jan,Deghi2020Jul,Ciliberto2020Nov,Sanchez2019Nov,Hajiloo2020Oct,Lu2021Feb,Gao2023Nov,Acciai2024Feb}.
In such systems, finite power production or even multiple useful outputs~\cite{Manzano2020Dec,Liu2020Apr,Lopez2023Jan,Lu2023Feb} can be achieved even when there is no \textit{average} heat exchange with the resource. However, nonvanishing fluctuations in the energy exchange between resource and working substance are required for the demonic operation of the device in this case.
A whole range of demonic devices spanning from the purely information-driven to nonthermal demons and beyond has been proposed~\cite{Krause2011Nov,Sanchez2019Oct,Fu2021Oct,Freitas2021Mar,Bozkurt2023Dec,Picatoste2024Aug}, where these devices have in common that they \textit{seemingly} violate the second law, see Ref.~\cite{Whitney2023Apr} for a detailed discussion.

In this paper, we analyze a triple-dot setup~\cite{Whitney2016Jan}, as shown in Fig.~\ref{fig:setup}, which acts as a demon exploiting both information \textit{and} a nonthermal resource.
Consequently, it is able to operate as an autonomous demon in the spirit of the devices of Refs.~\cite{Strasberg2013Jan,Koski2015Dec}, but with the advantage of having an exactly vanishing average heat flow between resource and working substance in a broad range of parameters and at non-vanishing output cooling power.
The idea underlying this device relies on a nontrivial combination of thermal machines based on capacitively coupled quantum dots, exploiting heat as a resource in order to implement a useful task~\cite{Sanchez2011Feb,Thierschmann2015Oct,Zhang2016Jan,Dare2017Sep,walldorf:2017,Bhandari2018Jul,dare:2019,Xi2021Sep,Tabatabaei2022Sep} and analogous information-based devices~\cite{Strasberg2013Jan,Ito2015Jun,Ptaszynski2018Jan,Erdman2018Jul,Poulsen2022Apr}, which can also be viewed as thermal machines exploiting a cold resource~\cite{Whitney2023Apr}.

The system we are studying in this paper is depicted in Fig.~\ref{fig:setup}. The goal of the shown engine is to extract heat from the colder reservoir of the working substance via electron transport through the central quantum dot W. The resource region, which we, on equal terms, refer to as demon, consists of two quantum dots, capacitively coupled to each other and to the working-substance dot W, and tunnel-coupled to one hot and one cold resource reservoir. The basic mechanism is based on the tunneling rates of the system depending on its interaction with the demon: tunneling rates in the system are higher for the left contact when the demon dots are empty and from the right when they are occupied. Henceforth charge fluctuations in the demon modify not only the system energy but also its symmetry. In this way, trajectories in which an electron goes through the system upon exchanging energy with the demon (via the Coulomb interaction) have a preferred direction. Note that while this combination of hot and cold resources might seem a very specific choice it can to a large extent be representative of \textit{any} nonthermal resource since only specific points in the energy distributions of the resource reservoirs are probed via the addition energies of the resource dots H and C.

Importantly, while the system works as a refrigerator, we impose a fully vanishing average energy (or heat) flow between the working substance and the resource region, which hence occurs to be ``demonic''. We analyze the cooling power, as well as entropy production and information exchange between the two parts of this bipartite system, using a steady-state master equation approach, and an unraveling in state-space trajectories.
In addition, we use full counting statistics
(FCS)~\cite{nazarov_quantum_2003,Bagrets2003Feb,Esposito2009Dec} in order to reveal the precision of the engine. While this triple-dot system has previously been analyzed using a master equation approach~\cite{Whitney2016Jan} focusing mostly on energy exchange, the FCS analysis~\cite{Sanchez2012Nov,Sanchez2013Dec}, the trajectory unraveling~\cite{Mayrhofer2021Feb,Wang2022Feb}, and the study of average information flows~\cite{Horowitz2014Jul, Shiraishi2015Jan, Ptaszynski2019Apr, Schreiber2000Jul, Takaki2022Nov} have until now been mostly limited to double-dot systems working either as heat engines or as information-driven engines exploiting a cold resource.
In particular, a study of information flows at the trajectory level has been fully missing and is provided in the present paper, and it here appears as a valuable tool to understand the role of the different resources in the thermodynamic performance of the device.
Indeed, charge detection based FCS experiments allow for the detection of trajectories in the state space of quantum dots~\cite{Schleser2004Sep,Vandersypen2004Nov,Gustavsson2006Feb,Fujisawa2006Jun,Hofmann2016Nov}, giving access to the thermodynamic flows~\cite{Sanchez2012Nov,Koski2013Oct,Koski2014Sep,Hofmann2016Jan,Hofmann2017Mar,Chida2022Oct}, fluctuations~\cite{Kung2012Jan,Saira2012Oct} and even mutual information~\cite{Koski2014Jul}.

It is the combination of steady-state master-equation analysis, FCS, and thermodynamic- and information-analysis at the trajectory level which allows us to identify two different operation regimes of the device. These two regimes lead to similar cooling power and efficiency, but to very different performance in view of precision. Importantly, we find one operation regime where both the interaction with the cold and the hot resource reservoir are beneficial for the cooling process, while only the interaction with the cold reservoir exploits information. This working principle leads to cooling at good precision. By contrast, a more information-focused operation process results in a competition between cooling and heating in the working substance and thereby comes at the cost of reduced precision.

This paper is structured as follows: We introduce the general framework in Sec.~\ref{sec:framework}, where we give details about the studied model (Sec.~\ref{sec:model}) the master equation formalism both for average currents as well as for the cooling-power noise from full counting statistics (Sec.~\ref{sec:average}) and the trajectory analysis of the model (Sec.~\ref{sec:traj}). In Sec.~\ref{sec:case}, we analyze the performance of the triple-dot based bipartite information-thermal machine. We demonstrate how the state-space cycles underlying the operation principle of the device enter the currents of interest in Sec.~\ref{sec:currents-cycles}, and analyze parameter regimes of optimal performance with respect to cooling power and precision in Sec.~\ref{sec:average performance}. Differences in the trade-off between cooling power and precision are analyzed based on the operational principle in Sec.~\ref{sec:discussion}. We conclude in Sec.~\ref{sec:conclusions}. Some analytical details and additional numerical results are presented in the Appendix.

In the following, we use the convention $\hbar,\kB  \equiv 1$.

\section{General framework}\label{sec:framework}

Here, we present the general framework of thermoelectric effects and Coulomb interaction in the triple quantum-dot system. Notably, experimental control of quantum dot systems in the Coulomb blockade regime~\cite{ihn_semiconductor_2009} has enabled useful operations as thermoelectric engines~\cite{staring_coulomb_1993,dzurak_observation_1993,dzurak_thermoelectric_1997,Scheibner2007Jan,svensson_lineshape_2012,thierschmann_diffusion_2013,svensson_nonlinear_2013,Thierschmann2015Oct,josefsson_quantum_2018,Prete2019May} and for single-electron heat control~\cite{thierschmann_thermal_2015,dutta_thermal_2017,dutta_single_2020}, also exploiting nonequilibrium phonons~\cite{dorsch:2020,Kuroyama2022Aug}, photons~\cite{Asgari2021Oct,Haldar2024Feb} or noise~\cite{Chida2023May}.

In this paper, we study a setup made of three capacitively coupled quantum dots, as depicted in Fig.~\subfigref{fig:setup}{a}.

\subsection{Triple-dot setup}\label{sec:model}

For each dot, we consider a single level with energy $\epsilon_i$, where $i=\H,\M,\C$. We assume that the intradot Coulomb interaction is large enough to exclude double occupation of a single dot. While the spin-degeneracy is known in general to influence the decay rate of the quantum-dot energy~\cite{Schulenborg2016Feb} and energy transport~\cite{Schulenborg2017Dec}, this is not of qualitative relevance for the effect observed here and we therefore neglect the spin degree of freedom of the electrons in the following.
The Hamiltonian describing the three quantum dots is therefore given by
\begin{equation}\label{Hdot}
    \hat{H}=\sum_{i=\H,\M,\C}\epsilon_i\hat{n}_i+\UH\hat{n}_\H\hat{n}_\M+\UC\hat{n}_\C\hat{n}_\M+U\hat{n}_\H\hat{n}_\C,
\end{equation}
where we have denoted the electron number operator in dot $i$ by $\hat{n}_i$. The interdot interaction energies are given by $U_\mathrm{H},U_\mathrm{C}$ and $U$. Here, we assume for simplicity that the interaction energy between the two resource dots, H and C, see Fig.~\subfigref{fig:setup}{a}, is very large with respect to temperatures, $U/\text{max}_\alpha(T_\alpha) \to \infty$, so that simultaneous occupation of these dots is excluded. A finite interaction energy between dots H and C was considered in Ref.~\cite{Whitney2016Jan} and was shown to merely have a quantitative effect, while not hindering the operation of the demon.

Each resource dot is furthermore tunnel-coupled to a single terminal, that is a large fermionic reservoir characterized by its temperature $T_\alpha$ and chemical potential $\mu_\alpha$, with $\alpha = \C,\H$. We choose $T_\C < T_\H$ and therefore call the corresponding terminals the cold and hot resource reservoirs. Since we here mostly focus on situations where the resource does not provide any energy on average, we refer to the resource region both as ``resource'' or ``demon''.
The working-substance dot W, in the lower part of the setup, is tunnel-coupled to two terminals, called the left (L) and right (R) reservoirs, with respective temperatures $T_\L$, $T_\R$ and chemical potentials $\mu_\L$, $\mu_\R$.
The condition of zero energy exchange between resource and working substance is achieved by fixing one of the model parameters, see Sec.~\ref{sec:average performance} for details.

For simplicity, we assume that the tunnel rates $\GC$ and $\GH$ of the upper terminals are energy independent.
However, it is crucial in this work to have energy-dependent tunnel rates for the left and right reservoirs. Given that the energy at which the electrons tunnel in and out of the dot W is fully determined by the occupations $h$ and $c$ of the upper dots, we denote the lower tunnel rates $\Gamma_{\L/\R}^{hc}$. Note that this is a property of the system imposed by the Coulomb interaction of the electrons in the different dots, not interpreted as a backaction effect of the demon acting on the barriers~\cite{Sanchez2019Oct}.

Since the only possibly relevant potential \textit{difference} in this setup is $\mu_\mathrm{L}-\mu_\mathrm{R}$, we here set the energy references such that $\mu_\L = \mu_\C = \mu_\H = 0$, without loss of generality.

\subsection{Steady-state dynamics}\label{sec:average}

\subsubsection{Master equation}\label{sec:master}

In the following, we consider the weak coupling (sequential tunneling) limit, i.e., $\Gamma_\alpha \ll T_\alpha$ with $\Gamma_\alpha$ the coupling strength (tunnel rate) between reservoir $\alpha$ and the nearest dot.
In this regime\footnote{Also note that coherences do not play a role for the dynamics here, since we consider spinless and purely capacitively coupled quantum dots.}, higher-order tunneling processes are negligible and we can describe the steady-state dynamics of the triple-dot system by a master equation approach \cite{Breuer2010}. We therefore introduce the  vector of occupation probabilities of triple-dot states
$\P = (p_{000}, p_{001}, p_{010}, p_{011}, p_{100}, p_{110})^\text{T}$, where $p_{hwc}$ denotes the probability that the dots H, W, and C contain respectively $h$, $w$, and $c$ electrons, with $h,w,c\in\{0,1\}$.  We have left out $p_{101}$ and $p_{111}$ since we exclude the simultaneous occupation of both resource dots. Then, the dynamics of the system is given by the rate equation
\begin{equation}
    \difc{\P}{t} = W\P,     \label{eq:master}
\end{equation}
where the kernel $W$ is a $6\times6$ matrix which can be decomposed into contributions from each reservoir, $W = \sum_{\alpha=\L,\R,\H,\C} W_\alpha$ \cite{Whitney2016Jan}. In the occupation basis, the off-diagonal kernel elements are
\begin{align}
    [W_\C]_{hwc}^{h'w'c'} &= \delta_{h,h'}\delta_{w, w'}\delta_{c, 1 -c'}\GC f_\C(\Delta E_{hwc}^{h'w'c'}),\nonumber\\
    [W_\H]_{hwc}^{h'w'c'} &= \delta_{h,1-h'}\delta_{w, w'}\delta_{c,c'}\GH f_\H(\Delta E_{hwc}^{h'w'c'}),\label{W elts}\\\nonumber
    [W_{\L/\R}]_{hwc}^{h'w'c'} &= \delta_{h,h'}\delta_{w,1-w'}\delta_{c, c'}\Gamma_{\L/\R}^{hc} f_{\L/\R}(\Delta E_{hwc}^{h'w'c'}),
\end{align}
where $f_\alpha(E) = 1/(1 + \e^{\beta_\alpha(E - \mu_\alpha)})$ is the Fermi-Dirac distribution and $\Delta E_{hwc}^{h'w'c'} = E_{hwc} - E_{h'w'c'}$ is the addition energy, namely the energy difference between the states $hwc$ and $h'w'c'$, with $E_{hwc}= c\eC + h\eH + w\eM + cw\UC + hw\UH$ [Eq.~\eqref{Hdot}]. Note that the sequential-tunneling approximation only allows for transitions in which the occupation of one of the dots changes by one.
The diagonal elements are given by $[W_\alpha]_{hwc}^{hwc} = -\sum_{h'w'c'\neq hwc} [W_\alpha]_{hwc}^{h'w'c'}$.

While this notation for the kernel is most general, we will in this paper also use the more intuitive notation indicating whether an electron enters ($+$) or leaves ($-$) a dot
\begin{eqnarray}
    &\GH^{w+}\equiv [W_\H]_{1wc}^{0wc},\ \quad \ &\GH^{w-}\equiv [W_\H]_{0wc}^{1wc},\nonumber\\
    &\GC^{w+}\equiv [W_\C]_{hw1}^{hw0},\ \quad \ &\GC^{w-}\equiv [W_\C]_{hw0}^{hw1},\label{Gamma elts}\\\nonumber
    &\Gamma_{\L/\R}^{hc+}\equiv [W_{\L/\R}]_{h1c}^{h0c},\ \quad \ &\Gamma_{\L/\R}^{hc-}\equiv [W_{\L/\R}]_{h0c}^{h1c}.
\end{eqnarray}
As indicated by the red crosses in Fig.~\ref{fig:setup}, we here choose a specific situation, in which the energy dependence of the tunnel rates in the working substance allows for transitions to/from the left reservoir only when the dots H and C are empty, namely $\GL^{00\pm}\neq0$ and $\GL^{10\pm}=\GL^{01\pm}=0$, while transitions to/from the right reservoir are allowed only if either dot H or dot C are occupied, namely $\GR^{00\pm}=0$ and $\GR^{10\pm},\GR^{01\pm}\neq0$. The working principle of the analyzed triple-dot demonic machine only requires an asymmetry in the energy-dependence of the coupling rates~\cite{Whitney2016Jan}, similar to the case of a standard heat engine implemented with a thermal resource~\cite{Rutten2009Dec,Sanchez2011Feb}; however, the above, fully asymmetric choice simplifies the analytic study significantly and thereby allows for novel insights into the role of heat and information as a resource.

In the following, we study the steady state of the device, denoted $\bP$, determined by $W\bP = 0$. This state is unique due to the kernel being regular and it can here be computed analytically, see Appendix~\ref{app:ss} for the full solution.

\subsubsection{Steady-state currents}
\label{sec:ss_currents}

Using the vector notations introduced above, the steady-state current for quantity $\nu$ (particle number $N$, energy $E$, heat $Q$, \dots) from reservoir $\alpha$ is, in the weak-tunneling regime, given by~\cite{Schulenborg2018},
\begin{equation}\label{average_current}
    J^\nu_\alpha = (\mathbf{x}_\alpha^\nu)^\text{T} W_\alpha \bP,
\end{equation}
where $\mathbf{x}_\alpha^\nu$ is a vector in the occupation basis (000, 001, 010, 011, 100, 110) with entries $[\mathbf{x}_\alpha^\nu]_{hwc}$.
In particular, for the particle-, energy-, and heat currents, we have
\begin{align}\label{x}
    \mathbf{x}_\alpha^N &= \mathbf{N}, & \mathbf{x}_\alpha^E &= \mathbf{E},&  \mathbf{x}_\alpha^Q &= \mathbf{E} -\mu_\alpha \mathbf{N},
\end{align}
with $[\mathbf{N}]_{hwc} = h + w + c$ the occupation number and $[\mathbf{E}]_{hwc} = E_{hwc}$ the energy of the dots in state $hwc$.

One of the main quantities of interest is the cooling power of the refrigerator $\Pcool \equiv \JR$, namely the heat current out of the cold reservoir of the working substance (here chosen to be the right one without loss of generality). At the same time, we typically require that the resource used for this task is ``demonic'', namely that the heat current flowing out of the resource region into the working substance $\Jin = \JC + \JH$ equals zero
\footnote{Here, the heat current flowing into the working substance $\Jin$ equals the heat current out of the resource. This is due to the absence of particle exchange between the two subsystems and due to the choice of zero potential difference within the working substance, $J^Q_\alpha = J^E_\alpha$,  together with energy conservation.}.

Note that in contrast to the particle current $J^N\equiv J^N_\L = -J^N_\R$, heat currents are in general not conserved.
Instead, we have the total entropy production rate given by
\begin{equation}
    \dot{\Sigma} = -\sum_{\alpha=\L,\R,\H,\C} \beta_\alpha J^Q_\alpha \ge 0 \label{entropy production rate}
\end{equation}
when considering the full triple-dot system in its steady state given by $\bP$. The entropy production rate is constrained by the second law of thermodynamics; it allows us to define a meaningful efficiency, even in the case where the total heat current from the resource vanishes~\cite{Hajiloo2020Oct,Tesser2023Apr,Manzano2020Dec},
\begin{equation}\label{eq:global_eff}
    \etaG=\frac{\beta_\L J_\L^Q+\beta_\R J_\R^Q}{-\beta_\H J_\H^Q-\beta_\C J_\C^Q},
\end{equation}
in terms of the entropy production rate of the demon used as a resource.
We emphasize the fact that this efficiency definition is based on the global entropy production.

The components of the sum in Eq.~\eqref{entropy production rate}, $-\beta_\alpha J^Q_\alpha$, equal the entropy current, defined by~\cite{Ptaszynski2019Apr}
\begin{equation}\label{J^S}
    J^S_\alpha = (\log\bP)^\text{T}W_\alpha \bP,
\end{equation}
with $[\log\bP]_{hwc} = \log \p_{hwc}$, only in the special case where the reservoirs are thermal and the capacitive coupling between the quantum dots is zero. Instead, in the presence of capacitive coupling, the flow of information between dots contributes to the entropy production and has to be taken into account. We describe how to do this in the following.

\subsubsection{Information flow}\label{sec:average info flow}

We now split the system into a lower part (W), the working substance, and an upper part (D), the demon, see Fig.~\subfigref{fig:setup}{a}, where the working substance is shown with a light-green background. While we are interested in a situation where there is no heat or energy exchange between these two subsystems on average, the two systems are however never independent of each other due to the capacitive coupling between the dots. This dependence of the two components of this bipartite system can be cast in terms of \textit{mutual information}.
The mutual information between W and D reads
\begin{equation}
    \cI_{\M:\D} = \sum_{h,w,c} \p_{hwc} \log\left(\frac{\p_{hwc}}{\p^\M_{w}\p^\D_{h  c}}\right),
    \label{eq:mutual_info_def}
\end{equation}
with $\p^\M_{w} = \sum_{h,  c} \p_{hwc}$ and $\p^\D_{h c} = \sum_{w} \p_{hwc}$. We emphasize that as soon as there is interdot Coulomb interaction and W and D are out of equilibrium with respect to each other, there is mutual information between the working substance and the resource---even in cases where there is clearly no feedback involved in the refrigeration or work extraction process
\footnote{For example, infinitely hot reservoirs in the demon make the occupations of H and C fully independent of W, while the occupation of W still depends on the dynamics of H and C.}.

Here, we consider the steady state of the system, such that $\difc{\cI_{\M:\D}}{t} = 0$.
Nevertheless, the rate of change of the mutual information can be split into information flows into the working substance and demon respectively \cite{Ptaszynski2019Apr}, $\difc{\cI_{\M:\D}}{t} = J^I_\M+ J^I_\D = 0$, with
\begin{equation}
    J^I_\M = -J^S_\L - J^S_\R, \quad J^I_\D = -J^S_\C- J^S_\H,
\end{equation}
and the entropy currents $J^S_\alpha$ defined in Eq.~\eqref{J^S}. In the following, we denote $J^I \equiv J^I_\D = -J^I_\M$, such that a positive information flow, $J^I > 0$, means that the demon is receiving information about the occupation of dot W. Similarly, we split the entropy production rate from Eq.~\eqref{entropy production rate} into local entropy production rates in W and D~\cite{Ptaszynski2019Apr}, $\dot{\Sigma} = \dot{\Sigma}_\M+\dot{\Sigma}_\D$.
Both of them obey a local version of the second law,
\begin{subequations}\label{local second law}
\begin{align}
    \dot{\Sigma}_\M &= -\bL\JL - \bR\JR + J^I \ge 0,\label{second law M}\\
    \dot{\Sigma}_\D &= -\bC\JC - \bH\JH - J^I\ge 0.\label{second law D}
\end{align}
\end{subequations}

In the following, we refer to the relation $\Jin = 0$ (where $\Jin = \JC + \JH$ is the heat flow from D to W) as the demon condition. Indeed, for instance, at zero voltage bias and $\Jin = 0$, we also have $\JL = -\JR$ so we can easily see that when the setup operates as a refrigerator, i.e. $\JR > 0$, then we have $-\bL\JL - \bR\JR < 0$.
At first glance, this makes it seem that the second law is not fulfilled when considering only W, since $\Jin = 0$. But this reasoning forgets to account for the information flow between the two parts, as can be seen in Eq.~\eqref{second law M}.

In order to highlight the role of the information flow in the working principle of this thermal machine, one can define an \textit{information efficiency}, as an alternative to the global efficiency defined in Eq.~\eqref{eq:global_eff}. Using Eq.~\eqref{second law M}, we define
\begin{equation}\label{eq:info_eff}
    \etaI=\frac{\beta_\L J_\L^Q+\beta_\R J_\R^Q}{J^I}.
\end{equation}
Due to the local second laws of Eq.~\eqref{local second law}, one finds that in general
\begin{equation}
    1 \ge\etaI\geq\etaG,
\end{equation}
since the information efficiency does not account for effects of the local entropy production as it could in general arise from mechanisms like local Joule heating or Fourier heat conduction from the hot to the cold part of the resource.

\subsubsection{Cooling-power fluctuations from full counting statistics}\label{sec:noise}

A main quantity of interest in this paper, in addition to the average currents in the device, is the precision of the produced cooling power, or, in other words, the absence of cooling power fluctuations. We compute these fluctuations using full counting statistics (FCS).

We here consider the case $\mu_\R= 0$, such that the cooling power $\Pcool = \JR$ is equal to the energy current out of the right reservoir, $J_\R^E$. The (zero-frequency) noise in the cooling power is therefore given by $\Scool = \diff{\cum{E_\R^2}}{t}$ where $\cum{E_\R^2}$ is the variance (second cumulant) of the energy $E_\R$ transferred from the right reservoir into dot W. Similarly, the energy current defined in Eq.~\eqref{average_current} corresponds to the derivative of the first cumulant, $J^E_\R = \diff{\cum{E_\R}}{t}$.

In order to calculate this variance, we define a counting field $\xi$ for the energy flowing out of the right reservoir. Then, the master equation \eqref{eq:master} including the counting fields becomes~\cite{Sanchez2013Dec, Schaller2014Jan}
\begin{equation}\label{meq_FCS}
    \difcp{\P}{t}(t, \xi) = W(\xi) \P(t, \xi),
\end{equation}
with $ W(\xi) = W_\C + W_\H + W_\L + W_\R(\xi)$ and $[W_{\R}(\xi)]_{hwc}^{h'w'c'} = [W_{\R}]_{hwc}^{h'w'c'}\e^{i\Delta E_{hwc}^{h'w'c'}\xi}$ (see Eq.~\eqref{W elts}).
From this, one finds the moment-generating function as $\mathcal{G}(t, \xi) = \sum_{hwc}[\P(t, \xi)]_{hwc}$.
As the next step, the $k$-th cumulant of the energy is obtained by differentiating the cumulant-generating function $\mathcal{F}(t, \xi) = \log(\mathcal{G}(t, \xi))$,
\begin{align}\label{cumulants}
    \cum{E_\R^k}(t) &= (-i\partial_{\xi})^k \mathcal{F}(t, \xi)|_{\xi=0}.
\end{align}
In the long-time limit (of interest for our steady-state analysis), the cumulant-generating function can be approximated by $\mathcal{F}(t, \xi) \simeq \lambda(\xi)t$ \cite{Schaller2014Jan}, where $\lambda(\xi)$ is the eigenvalue of $W(\xi)$ such that $\lambda(\xi) \rightarrow 0$ for $\xi \to 0$ while all the other eigenvalues have negative real parts in this limit, due to the uniqueness of the steady state in the case of interest here.
Then, the noise in the cooling power is given by
\begin{equation}
    \Scool = -\partial^2_\xi{\lambda(\xi)}|_{\xi=0}\ .
\end{equation}
More precisely, we do the following to obtain the cumulants analytically. Starting from the moment-generating function in the long-time limit \cite{Sanchez2007Apr, Sanchez2013Dec}, we take the Laplace transform,
\begin{equation}\label{G(z, xi)}
    \mathcal{G}(z, \xi) = \sum_{hwc}\left[\frac{1}{z\mathds{1} - W(\xi)}\bar{\P}\right]_{hwc},
\end{equation}
where $\bar{\P}$ is the steady state of the master equation \eqref{eq:master} (i.e. for $\xi =0$). Then, the eigenvalue $\lambda(\xi)$ is the pole $z_0$ of $\mathcal{G}(z,\xi)$ close to $\xi = 0$.
Using the Taylor expansion of $z_0$ around $\xi = 0$,
\begin{equation}\label{z0}
    z_0 = \sum_{n \in\mathbb{N}} c_{n} \frac{(i\xi)^n}{n!},
\end{equation}
in the denominator of $\mathcal{G}(z, \xi)$ [Eq.~\eqref{G(z, xi)}], we determine the coefficients $c_{n}$ by proceeding order by order. In this way, one obtains $\Pcool = c_1$ and $\Scool = c_2$.

In order to analyze the performance accounting for the trade-off between cooling power, efficiency, and precision, we define the performance quantifier
\begin{equation}\label{X_TUR}
    X_\mathrm{TUR,global/info} = 2 \Pcool\frac{\eta_\mathrm{global/info}}{1-\eta_\mathrm{global/info}}\frac{\TL}{\Scool}\eta_\text{Carnot}
\end{equation}
inspired by the thermodynamic uncertainty relations (TUR)~\cite{Pietzonka2018May,Kheradsoud2019Aug,Gingrich2016Mar} see explanations in Appendix~\ref{app:TUR}. We have denoted $\eta_\text{Carnot} = \TR/(\TL - \TR)$ the Carnot efficiency of a standard refrigerator operating between temperatures $\TR$ and $\TL$.
In particular, under demon condition, $J_\mathrm{in}^Q=0$, and at zero potential bias, $\mu_\R = 0$, $X_\mathrm{TUR,global}$ corresponds exactly to the TUR for the three-dot system~\cite{Gingrich2016Mar}. Therefore $X_\mathrm{TUR,global}$ is bounded by 1, like the standard TUR for refrigerators, however, we cannot make a similar statement about $X_\mathrm{TUR,info}$, see details in Appendix~\ref{app:TUR}.
Instead, since $\dot{\Sigma} = \dot{\Sigma}_\M + \dot{\Sigma}_\D$, from the local second laws \eqref{local second law}, and using the expressions of the efficiencies, given in Eqs.~\eqref{eq:global_eff} and \eqref{eq:info_eff}, we find that in general
\begin{equation}
    X_\mathrm{TUR,info} \ge X_\mathrm{TUR,global}.
\end{equation}
In Sec.~\ref{sec:average performance}, we will use both quantifiers to analyze the performance of the triple-dot system operating as a ``demonic'' refrigerator.

\subsection{Trajectory analysis of the dynamics}\label{sec:traj}

To get further insight into how the device operates, it is interesting to look at the trajectories corresponding to the Markovian stochastic process described by the master equation \eqref{eq:master}. Indeed, we can define the thermodynamic quantities we are interested in at the trajectory level. Furthermore, we can use our insight on the stochastic dynamics to tune the parameters of the system to favor trajectories with useful outcomes ---here specifically transporting heat out of the right reservoir.

\subsubsection{Trajectories for the entire triple-dot system}

We consider trajectories in the triple-dot state space $\gamma = (\gamma_0, ..., \gamma_M)$, where $\gamma_m = h_m w_m c_m$ is the state of the dots at time $t_m = m\delta t$, and $\tilde{\gamma}$ denotes the time-reversed trajectory $(\gamma_M, ..., \gamma_0)$.
We have discretized time with a time step $\delta t$ such that $\tau_\text{th} \ll \delta t \ll 1/\max_\alpha\Gamma_\alpha$, where $\tau_\text{th}$ is the characteristic thermalization time of the reservoirs \cite[Chapter 4]{Haroche}.

The probability of trajectory $\gamma$ is given by
\begin{equation}
    P(\gamma) = \p({\gamma_0}) \prod_{m=1}^{M} P_{\alpha_m}[\gamma_{m}|\gamma_{m-1}],
\end{equation}
with the steady-state probability $\p(\gamma_0) = \p_{h_0 w_0 c_0}$ of finding the system in the state $h_0 w_0 c_0$ at $t=0$ and $P_{\alpha_m}[\gamma_{m}|\gamma_{m-1}]$ the probability of the transition $\gamma_{m-1} \xrightarrow{\alpha_m} \gamma_m$, where $\alpha_m = \C,\H,\L,\R$ indicates which reservoir is responsible for the transition\footnote{Strictly, we only need the $\alpha_m$ notation for dot W since it is the only dot connected to two different reservoirs, which can induce transitions.}. The no-jump evolution ($\gamma_{m} = \gamma_{m-1})$ is denoted by $\alpha_m = 0$.
The transition probability is given by
\begin{equation}
    P_{\alpha_m}[\gamma_m|\gamma_{m-1}] = \left\{
    \begin{array}{ll}
        [W_{\alpha_m}]^{\gamma_{m-1}}_{\gamma_m}\delta t & \text{if $\gamma_m \neq \gamma_{m-1}$}\\
        (1 + [W]^{\gamma_m}_{\gamma_m}\delta t)\delta_{\alpha_m,0} & \text{if $\gamma_m = \gamma_{m-1}$}\\
    \end{array}\right. .
\end{equation}
With this definition, if the transition $\gamma_{m-1} \xrightarrow{\alpha_m} \gamma_m$ is not feasible in the setup (e.g. $\alpha_m = \C$ while changing the occupation of dot W), we clearly have $P_{\alpha_m}[\gamma_m|\gamma_{m-1}] = 0$. Furthermore, we have $P_{\alpha_m}[\gamma_m|\gamma_{m}] \le 1$ and, $\forall \gamma_{m-1}$, $\sum_{\gamma_m}\sum_{\alpha_m} P_{\alpha_m}[\gamma_m|\gamma_{m-1}] = 1$ since $[W]^{\gamma_m}_{\gamma_m}<0$ and $\sum_{\gamma_{m}} [W]^{\gamma_{m-1}}_{\gamma_{m}} = 0$, see Eq.~\eqref{W elts}.

At the single-trajectory level, the entropy production is given by~\cite{denBroeck2015}
\begin{equation}\label{eq:FR}
    \Sigma(\gamma) = \log\frac{P(\gamma)}{P(\tilde{\gamma})},
\end{equation}
where $P(\tilde\gamma)$ is the probability that the triple-dot system follows the time-reversed trajectory\footnote{Note that we do not need to modify the probability function to the probability of the time-reversed system because we are looking at steady state and constant external parameters.}, $\tilde\gamma$.
The entropy production can be split into $ \Sigma(\gamma) = \Delta S(\gamma) + \sum_\alpha \Delta S_\alpha(\gamma)$~\cite{Seifert2008Jan} where
\begin{align}
\label{eq:entropy_production_splitting}
    \Delta S(\gamma) &= \log\left(\frac{\p(\gamma_0)}{\p(\gamma_M)}\right), \\\nonumber
    \Delta S_\alpha(\gamma) &= \sum_{m=1}^M\delta_{\alpha_m \alpha}\log\left(\frac{P_{\alpha}[\gamma_{m}|\gamma_{m-1}]}{P_{\alpha}[\gamma_{m-1}|\gamma_{m}]}\right) = -\beta_\alpha Q_\alpha(\gamma),
\end{align}
are respectively the stochastic entropy variation of the dots and of reservoir $\alpha$, with $Q_\alpha(\gamma)$ the heat received by the system from reservoir $\alpha$ along the trajectory $\gamma$.
One recovers all quantities defined in Sec.~\ref{sec:average info flow} by averaging over all possible trajectories and dividing by the duration, e.g., $\dot{\Sigma} = \lim_{M\to \infty}\mean{\Sigma(\gamma)}_{\gamma}/t_M$.

\subsubsection{Trajectories for the bipartite system: demon and working substance}

In the same spirit as in Sec.~\ref{sec:average info flow}, we now define the stochastic thermodynamic quantities for a single subsystem of the bipartite systems, consisting either of the working substance W or the (demonic) resource D. Following Refs.~\cite{Horowitz2014Jul, Takaki2022Nov}, we define the subsystem entropy variations
\begin{gather}
\begin{aligned}
    \Delta S_\M(\gamma) &= \log\left(\frac{\p^\M_{w_0}}{\p^\M_{w_N}}\right),\\
    \Delta S_\D(\gamma) &= \log\left(\frac{\p^\D_{h_0 c_0}}{\p^\D_{h_N c_N}}\right),
\end{aligned}
\end{gather}
(see below Eq.~\eqref{eq:mutual_info_def} for the definition of the probabilities inside the logarithms) and the information acquired by the working substance and demon respectively,
\begin{gather}
\begin{aligned}
    \cI_\M(\gamma) &= \log\left(\prod_{m=1}^M\frac{\p(h_{m-1},c_{m-1}| w_{m})}{\p(h_{m-1},c_{m-1}| w_{m-1})}\right),\\
    \cI_\D(\gamma) &= \log\left(\prod_{m=1}^M\frac{\p(w_{m-1}|h_{m},c_{m})}{\p(w_{m-1}| h_{m-1},c_{m-1})}\right).\label{info traj}
\end{aligned}
\end{gather}
Here, $\p(w|h,c)$ denotes the (steady-state) conditional probability of finding occupation $w$ in W, knowing that the demon dots are in state $(h, c)$ and vice-versa for $\p(h,c|w)$.

With these definitions, we can  write the local entropy productions of the two subsystems
\begin{gather}
\begin{aligned}
    \Sigma_\M(\gamma) &= \Delta S_\M(\gamma) - \bL Q_\L(\gamma) - \bR Q_\R(\gamma) - \cI_\M(\gamma),\\
    \Sigma_\D(\gamma) &= \Delta S_\D(\gamma) - \bC Q_\C(\gamma) - \bH Q_\H(\gamma) - \cI_\D(\gamma).\label{stoc loc entropy prod}
\end{aligned}
\end{gather}
By averaging over all possible trajectories, we recover the local second laws given in Eq.~\eqref{local second law} from Sec.~\ref{sec:average info flow}. Note that we can also define trajectory-resolved quantities corresponding to the mutual information and entropy currents, see Appendix \ref{app:extra info qties trajectory}.

\section{Triple-dot refrigerator}\label{sec:case}

We are interested in the refrigerator operating mode of the device shown in Fig.~\ref{fig:setup}. We therefore assume that $T_\R < T_\L$ and define $\T = (T_\L + T_\R)/2$ and $\dT = T_\L - T_\R >0$. We recall that the cooling power is then directly given by the heat flow out of the right reservoir, $\Pcool \equiv \JR$. Furthermore, we assume that all electrochemical potentials are equal, in particular $\mu_\L=\mu_\R$, since we are not focusing on possible thermoelectric effects (e.g., Peltier cooling) which might otherwise occur in the working substance independently of the demonic resource.

\subsection{Operation cycles acting in parallel}\label{sec:fridge}

From now on, we focus on the special case depicted in Fig.~\subfigref{fig:setup}{a}, in which trajectories in state space can occur following the scheme of Fig.~\subfigref{fig:setup}{b}.
Concretely, this means that we only allow for a certain set of tunneling transitions, which is selected by a fully asymmetric energy dependence of tunnel rates. We assume that tunneling from and to dot W can occur from the right reservoir only when one of the dots H and C in the resource region is occupied, namely $\GR^{00} = 0$ and $\GR^{10},\GR^{01} \neq 0$. At the same time, we assume that tunneling from and to dot W due to the left reservoir can occur only when dots H and C are empty, namely $\GL^{10} = \GL^{01}= 0$ and $\GL^{00}\neq0$. The non-vanishing transition rates are found from Fermi's golden rule and given by
\begin{gather}
\begin{aligned}
    \Gamma^{w\pm}_{\C/\H} &= \Gamma_{\C/\H} f_{\C/\H}\bm{(}\pm(\epsilon_{\C/\H} + w U_{\C/\H})\bm{)},\\
    \Gamma^{hc\pm}_{\L/\R} &= \Gamma_{\L/\R}^{hc} f_{\L/\R}\bm{(}\pm(\eM + h\UH + c\UC)\bm{)},
\end{aligned}
\end{gather}
see also Eqs.~\eqref{W elts} and \eqref{Gamma elts}.
Here, we have furthermore used that the Coulomb interaction between dot H and C is infinitely large, meaning that simultaneous occupation of the two resource dots is excluded. This assumption implies that heat exchange between reservoirs H and C can occur only via dot W and not directly within the resource region, i.e., Fourier heat flow is weak within the resource region.

This fact together with the tunnel-rate asymmetry described in this section, implies that the operation of the triple-dot device is largely based on heat- and information exchange with the two dots H and C---and hence with the two reservoirs H and C---separately. Therefore, the device exhibits two operation cycles involving a single resource reservoir acting {\it in parallel} to achieve cooling in the absence of average heat flow between the working substance and the resource region, as will be explained in the next section, Sec.~\ref{sec:currents-cycles}.

This situation is in contrast to Ref.~\cite{Whitney2016Jan}, where the nonequilibrium character of the combined resource region was explicitly apparent.
Here, we instead make a parameter choice which rather allows us to focus on the interplay of heat-driven~\cite{Sanchez2011Feb,Thierschmann2015Oct} and information-driven~\cite{Strasberg2013Jan,Erdman2018Jul} machines, enabling a demonic effect in the absence of \textit{any} average heat flow in a broad range of parameters.

\begin{figure}[tb]
    \includegraphics[width=0.75\linewidth]{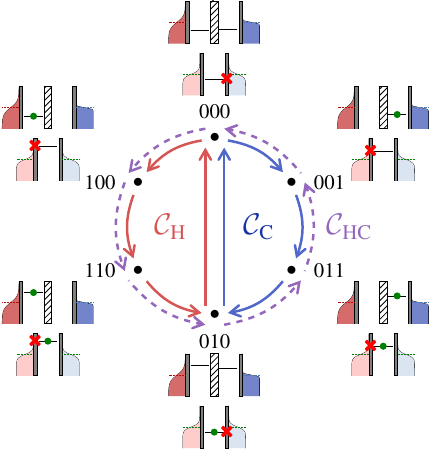}\vskip-3mm
    \caption{\label{fig:cycles}
        Graphical representation of the stochastic cycles from Table~\ref{tab:cycles}, $\CC$ (solid blue), $\CH$ (solid red) and $\CHC$ (dashed purple). See Fig.~\subfigref{fig:setup}{b} for the transition rates associated with each arrow. Green dots indicate the presence of electrons in the dots and red crosses correspond to blocked transitions.
    }
\end{figure}

\begin{table*}[htb]
    \setlength{\tabcolsep}{5pt}
    \renewcommand{\arraystretch}{1.75}
    \begin{tabularx}{0.92\linewidth}{ccccccccc}
        \hline\hline
        Cycle & Jump sequence & $N$ \hfill\,& $Q_\C$ \,& $Q_\H$\, & $Q_\L$\, & $Q_\R$\, & $\Sigma$\,  & $\cI$\, \\
        \hline
        $\CC$    & $\gCC = \GCop\GRiop\GCim\GLoom$             & $-1$ & $-\UC$ & $0$    & $-\eM$ & $\eM {+} \UC$ & $\Delta\beta_{\C\R}\UC {-} \Delta\beta_{\R\L}\eM$  & $\log\left(\frac{\p_{001}\p_{010}}{\p_{000}\p_{011}}\right)$\\
        $\CH$    & $\gCH = \GHop\GRiop\GHim\GLoom$             & $-1$ & $0$    & $-\UH$ & $-\eM$ & $\eM {+} \UH$ & $-\Delta\beta_{\R\H}\UH {-} \Delta\beta_{\R\L}\eM$ & $\log\left(\frac{\p_{100}\p_{010}}{\p_{110}\p_{000}}\right)$
        \\
        $\CHC$  & $\gCHC = \GHop\GRiop\GHim\GCip\GRoim\GCom$  & $0$  & $\UC$  & $-\UH$ & $0$    & $\UH {-} \UC$ & $-\Delta\beta_{\C\R}\UC {-} \Delta\beta_{\R\H}\UH$  & $\log\left(\frac{\p_{100}\p_{011}}{\p_{110}\p_{001}}\right)$\\
         \hline\hline
    \end{tabularx}
    \caption{\label{tab:cycles}
        Cycles and associated exchanged heat and particles corresponding to the Markov chain represented in Fig.~\subfigref{fig:setup}{b}, with parameter choices indicated in Sec.~\ref{sec:fridge}. See also Fig.~\ref{fig:cycles} for a graphical representation of these cycles. The first two cycles, $\CC$ and $\CH$, make a fundamental set that allows one to construct all the other cycles [Eq.~\eqref{cycle}] and compute the associated thermodynamic quantities [Eq.~\eqref{X(C)}].
        From this table, we can easily identify which cycles contribute to the refrigeration, namely $Q_\R(\cC) > 0$, and  whether $\cC$ is more likely to occur than its time-reverse $\bcC$, namely $\Sigma(\cC) > 0$, for a given set of parameters. We have defined $\Delta \beta_{\alpha \alpha'} = \beta_\alpha - \beta_{\alpha'}$.
    }
\end{table*}

\subsection{Average currents from basic cycles}\label{sec:currents-cycles}

In this section, we establish links between the average currents and noises defined in Sec.~\ref{sec:average} and the stochastic version of the quantities, as defined in Sec.~\ref{sec:traj}.

\subsubsection{Basic cycles}\label{sec:basic cycles}

As a first step, we take a stochastic trajectory $\gamma$ and cut it, without loss of generality, every time the system goes back to state $000$. In this way, we get all possible cyclic jump sequences $\cC'$ starting from $000$ in the Markov chain dynamics of the triple-dot system, represented in Fig.~\subfigref{fig:setup}{b}. We now want to classify these cycles based on their thermodynamic properties. We compute the stochastic thermodynamic quantity $X(\cC')$ along $\cC'$, with $X = Q_\alpha, \Sigma, \Sigma_{\M/\D}, I_{\M/\D}$, using the definitions from Sec.~\ref{sec:traj}.
Given that self-retracing jump sequences such as $\GH^{w+}\GH^{w-}$ do not contribute to $X(\cC')$, then $X(\cC')=X(\cC)$ where the basic cycle $\cC \in \{\cC_0, \CC, \bCC, \CH, \bCH, \CHC, \bCHC\}$ is obtained by removing all the self-retracing parts from $\cC'$.
This basic cycle set is not the most fundamental one, as will be detailed in Sec.~\ref{sec:fundamental cycles}, but is insightful from the thermodynamic perspective as it classifies cycles based on how the system interacted with the different resource reservoirs, see also Ref.~\cite{Hegde2024Aug} where a similar strategy was used to classify cycles in a two-terminal engine.
$\cC_0$ denotes the empty jump sequence, namely  $\cC'$ was fully self-retracing, such that $X(\cC_0) = 0$. The cycle $\CC$ involves only the cold resource reservoir, the cycle $\CH$ involves only the hot resource reservoir and the cycle $\CHC$ involves both of them, see graphical representation in Fig.~\ref{fig:cycles}. Their jump sequences, consisting of four or six jumps, is indicated in the second column of Table~\ref{tab:cycles}. The time-reversed version of cycle $\cC$ is denoted by $\bcC$.

Then, for each basic cycle $\cC$, one can evaluate the information acquired by the demon, $\cI_\D(\cC) = - \cI_\M(\cC) \equiv \cI(\cC)$ [Eq.~\eqref{info traj}] as well as the local entropy productions $\Sigma_\M(\cC) = -\bL Q_\L(\cC) - \bR Q_\R(\cC) + \cI(\cC)$ and $\Sigma_\D(\cC) = -\bH Q_\H(\cC) - \bC Q_\C(\cC) - \cI(\cC)$, expressed in terms of information and heat exchanges [Eq.~\eqref{stoc loc entropy prod}].
All relevant thermodynamic quantities for the basic cycles $\CC$, $\CH$, and $\CHC$ are listed in Table~\ref{tab:cycles}.

\subsubsection{Average currents}

The basic cycles ${\cC}$, as defined in Sec.~\ref{sec:basic cycles}, provide a thermodynamically relevant way to classify any cycle $\cC'$.
Indeed, we find that, in the considered case, the average currents [Eq.~\eqref{average_current}] can be decomposed into a sum of contributions of each basic cycle $\cC$,
\begin{equation}\label{current}
    J^\nu_\alpha = \sum_{\cC =\CC,\CH,\CHC} \left[\nu_\alpha(\cC)\rC +  \nu_\alpha(\bcC)r_{\bcC}\right].
\end{equation}
Here, $\nu = N, E, Q, I, S$
and $\nu_\alpha(\cC)$ is the total transported quantity along cycle $\cC$ by the considered current. Equation~\eqref{current} can either be obtained by direct calculation, or by relying on methods from network theory~\cite{Schnakenberg1976Oct}, see Appendix~\ref{app:graphs} for details. When we analyze $\nu = N, E, Q, S$, the index $\alpha$ indicates reservoir C, H, L or R. When we analyze $\nu = I$ (i.e., the information flows), the index $\alpha$ indicates subsystem W or D. The cycle quantity $S_\alpha(\cC)$ for the entropy current [Eq.~\eqref{J^S}], $\nu = S$, is defined in Appendix \ref{app:extra info qties trajectory}; please note that it should not be confused with the entropy variation of reservoir $\alpha$, $\Delta S_\alpha(\cC)$, defined by Eq.~\eqref{eq:entropy_production_splitting}.
The basic cycles contribute to all average currents with weights $\rC$, which we express in terms of tunnel rates:
\begin{gather}
\begin{aligned}
    \rCC &= \frac{\gH\gCC}{\g},& \rCH &= \frac{\gC\gCH}{\g},&  \rCHC &= \frac{\gCHC}{\g},\\
    \rbCC &= \frac{\gH\gbCC}{\g},& \rbCH &= \frac{\gC\gbCH}{\g},& \rbCHC &= \frac{\gbCHC}{\g},\label{cycle rates}
\end{aligned}
\end{gather}
with
\begin{gather}
\begin{aligned}
    \gC &= \GCom\GCim +\GCom\GRoim  +\GCim\GRoip, \\
    \gH &= \GHom\GHim +\GHom\GRiom  +\GHim\GRiop .
\end{aligned}
\end{gather}
In these expressions, $\gCC$, $\gCH$, $\gCHC$ correspond to the jump sequences from Table \ref{tab:cycles}, and the corresponding quantities for the time-reversed cycles are obtained by reversing the jump sequences, namely swapping $+$ and $-$ in the transition rates. The denominator $\g$ is the normalization factor of the steady-state probabilities $\p_{hwc}$, as given in Appendix~\ref{app:ss}.

The cycle rates $r_\cC$ defined in Eq.~\eqref{cycle rates} can be linked to the probabilities $\pi(\cC)$ that a cycle $\cC'$ equivalent to basic cycle $\cC$ occurs, which are discussed in detail in Sec.~\ref{sec:discussion}.
We show this link starting from the expression of the average current in terms of trajectory average, $J^\nu_\alpha = \lim_{M\to \infty}\mean{\nu_\alpha(\gamma)}_{\gamma}/t_M$.
Since we are considering the long-time limit, we can neglect contributions to $\nu_\alpha(\gamma)$ from the small initial and final portions of $\gamma$ which are not part of a closed cycle, and write
\begin{align}\label{average from traj1}
    J^\nu_\alpha
    &=  \lim_{M\to \infty}\frac{1}{t_M}\sum_{\cC}\mean{\nu_\alpha(\cC) M_\cC(\gamma) + \nu_\alpha(\bcC) M_{\bcC}(\gamma)}_{\gamma}.
\end{align}
We have denoted $M_\cC(\gamma)$ the number of cycles $\cC'$ in trajectory $\gamma$ which are equivalent to basic cycle $\cC$ after removing the self-retracing parts. The cycle probability $\pi(\cC)$ tells us how often a cycle $\cC'$ equivalent to basic cycle $\cC$ occurs, therefore we can write $\mean{M_\cC(\gamma)}_{\gamma} = \pi(\cC)\mean{M_{\text{cycles}}(\gamma)}_{\gamma}$, where $M_{\text{cycles}}(\gamma)$ is the total number of cycles from $000$ back to $000$ of any kind in trajectory $\gamma$, including purely self-retracing ones. From Eq.~\eqref{average from traj1}, we thus get
\begin{align}\label{average from traj}
    J^\nu_\alpha
    &=  \sum_{\cC}\left[\nu_\alpha(\cC) \pi(\cC) + \nu_\alpha(\bcC) \pi(\bcC)\right]\lim_{M\to \infty}\frac{\mean{M_{\text{cycles}}(\gamma)}_{\gamma}}{t_M}\nonumber\\
    &=  \frac{1}{\bar{t}_\text{cycle}}\sum_{\cC}\left[\nu_\alpha(\cC) \pi(\cC) + \nu_\alpha(\bcC) \pi(\bcC)\right],
\end{align}
where the time $\bar{t}_\text{cycle}$ is the average duration of any cycle from $000$ to $000$, namely the average time it takes the system to come back to state $000$.
Identifying the terms in Eq.~\eqref{average from traj} with Eq.~\eqref{current}, we finally obtain
\begin{equation}\label{eq:raterelation}
    r_\cC = \frac{\pi(\cC)}{\bar{t}_\text{cycle}}.
\end{equation}

The average cycle time $\bar{t}_\text{cycle}$ is given by the recurrence time from state $000$. This recurrence time, in turn, is equal to the average holding time in $000$, namely $1/(\GC^{0+} +\GH^{0+} + \GL^{00+})$, divided by the steady-state occupation of state $000$ \cite{Anderson1991}. We therefore obtain the explicit expression for the average cycle time $\bar{t}_\text{cycle}$ in terms of tunnel rates and of the steady-state master equation solution as
\begin{equation}\label{t_cycle}
    \bar{t}_\text{cycle} = \frac{1}{\p_{000}(\GCop + \GHop + \GLoop)}.
\end{equation}
We find that the analytical expression of $\pi(\cC)$ obtained from Eqs.~\eqref{eq:raterelation} and \eqref{t_cycle} is in excellent agreement with our results from numerical simulations,  as we discuss in Secs.~\ref{sec:average performance} and \ref{sec:discussion}.

We now use the relation between rates $\rC$ and probabilities $\pi(\cC)$ given in Eq.~(\ref{eq:raterelation}) to simplify the general current expression and to thereby highlight the role of time-reversed cycles. Using furthermore the fact that $\nu_\alpha(\bcC) = -\nu_\alpha(\cC)$ and the fluctuation relation~\eqref{eq:FR}, we find
\begin{equation}\label{current2}
    J^\nu_\alpha= \sum_{\cC =\CC,\CH,\CHC} \nu_\alpha(\cC)\rC\left[1 - \e^{-\Sigma(\cC)}\right].
\end{equation}

\subsubsection{Average currents for cycles acting in parallel}\label{sec:fundamental cycles}
With these compact expressions for all currents of interest, we are now able to give a more precise meaning to the concept of ``cycles acting in parallel'' outlined Sec.~\ref{sec:fridge} by showing that all currents can be split into contributions where the working substance interacts separately with the cold and hot resource reservoirs. To that aim, we introduce the fundamental set of cycles $\{\CC, \CH\}$~\cite{Schnakenberg1976Oct} (see Appendix~\ref{app:graphs} for details). Any basic cycle $\cC$, as defined in Sec.~\ref{sec:basic cycles}, can be constructed from this fundamental set as
\begin{equation}\label{cycle}
    \cC = \sigma_\C \CC \oplus \sigma_\H \CH,
\end{equation}
with $\sigma_\alpha=-1,0,1$ and $\alpha = \C, \H$. The symbol $\oplus$ means the union of the two cycles where any transition they have in common has been removed \cite{Schnakenberg1976Oct}, for instance, $\bCC = -\CC$ or $\CHC = -\CC \oplus \CH$.
This decomposition allows us to express the generic thermodynamic quantity $X(\cC)$ in terms of contributions of the cold and hot cycles,
\begin{equation}\label{X(C)}
    X(\cC) = \sigma_\C X(\CC) + \sigma_\H X(\CH),
\end{equation}
with, in particular, $X(\CHC) = X(\CH) - X(\CC)$ and $X(\bcC) = -X(\cC)$. Using the decomposition \eqref{cycle} for each basic cycle in Eq.~\eqref{current}, we can write
\begin{align}\label{current split}
    J^\nu_\alpha &= \nu_\alpha(\CC)\FC +  \nu_\alpha(\CH)\FH,
\end{align}
where we have defined the fluxes
\begin{gather}
\begin{aligned}
    \FC =\,& \rCC + \rbCHC - \rbCC - \rCHC =- \frac{\JC}{\UC} ,\\
    \FH =\,& \rCH + \rCHC - \rbCH - \rbCHC =- \frac{\JH}{\UH}.\label{F(C)}
\end{aligned}
\end{gather}
This means that one can rewrite any average current in terms of the heat currents out of the hot and cold resource reservoirs. Equation~\eqref{current split} is particularly insightful for $\JR$ since it allows splitting the cooling power into $\Pcool = \Pcool^{\C} + \Pcool^{\H}$, with
\begin{equation}\label{Pcool^Ca}
    \Pcool^{\alpha} = Q_\R(\cC_\alpha)F^{\alpha},
\end{equation}
such that we can see whether reservoir $\alpha = \C,\H$ from the demon part contributes to the cooling, $\Pcool^{\alpha} > 0$, or is detrimental to it, $\Pcool^{\alpha} < 0$.
For the particle current, equation~\eqref{current split} becomes
\begin{equation}
    J^N =  \frac{\JC}{\UC} + \frac{\JH}{\UH}.
\end{equation}
This coincides with the result from Ref.~\cite[Eq.~(13)]{Whitney2016Jan}, which was obtained, unlike here, in the absence of temperature and voltage biases between the left and right reservoirs but for possibly nonzero tunnel rates $\GL^{10}$, $\GL^{01}$ and $\GR^{00}$.

We can also write the heat current out of the right reservoir, namely the cooling power, as
\begin{equation}\label{JR2}
    \JR= -(\eM + \bar{U})J^N -\delta U \left(\frac{\JH}{\UH} - \frac{\JC}{\UC}\right),
\end{equation}
with $\bar{U} = (\UH +  \UC)/2$ and $\delta U = (\UH-\UC)/2$. Interestingly, the heat current out of the left reservoir
\begin{equation}\label{JL2}
    \JL = \eM J^N
\end{equation}
is always proportional to the particle current, also referred to as ``tightly coupled''~\cite{Edwards1995Aug,Humphrey2002Aug,Esposito2009Apr}, due to the choice of $\Gamma^{10\pm}_\L=\Gamma^{01\pm}_\L=0$ (electrons exchanged between contact L and dot W all have the same energy, $\epsilon_\mathrm{W}$). This is in contrast to the cooling power, which---in addition to a tightly coupled component---also has a term proportional to $\delta U$ that is independent of the particle current.

Since we are specifically interested in the performances of the device operating under the demon condition, $\Jin = 0$, it is interesting to express this condition as a condition for the interaction energies in terms of the fluxes of the fundamental cycles. We find
\begin{equation}\label{demon condition}
    \frac{\UH}{\UC}
    = -\frac{\FC}{\FH}.
\end{equation}
While this is an insightful relation, showing how the parameter choice needs to connect to the occurrence rates, see Eq.~\eqref{F(C)}, in the following, we fix the demon condition by a choice of single-level energies (which are here implicit in the fluxes on the right-hand side of the equation).

From Eqs.~\eqref{JR2} and \eqref{JL2}, we get further insights into the interaction energies required for a refrigerator working under demon conditions. We note that imposing the demon condition $J_\mathrm{in}^Q=\JR+\JL=0$ for $\delta U = 0$, $\bar{U} \neq 0$ necessarily results in a vanishing average particle current in the working substance, $J^N = 0$. This means that the cooling power vanishes in that case, as can also be seen in Fig.~\ref{fig:maxPcool_UH} at $\UH/\UC = 1$.

\subsection{Steady-state  performance and precision}\label{sec:average performance}

We start by presenting the performance of the device working as a refrigerator while exploiting a resource in the absence of heat extraction from the resource, based on the steady-state transport properties presented in Sec.~\ref{sec:average}. This means that from now on, we strictly impose demon condition, namely the average heat (energy) flow from the resource to the working substance is zero, $\Jin=0$.
This can be achieved in different ways using some demon parameters as knobs, while the others are fixed. We have here decided to adjust the single-level energy of dot H, $\eH$,  which can easily be tuned experimentally using gate voltages.

\begin{figure}[tb]
    \includegraphics[width=\linewidth]{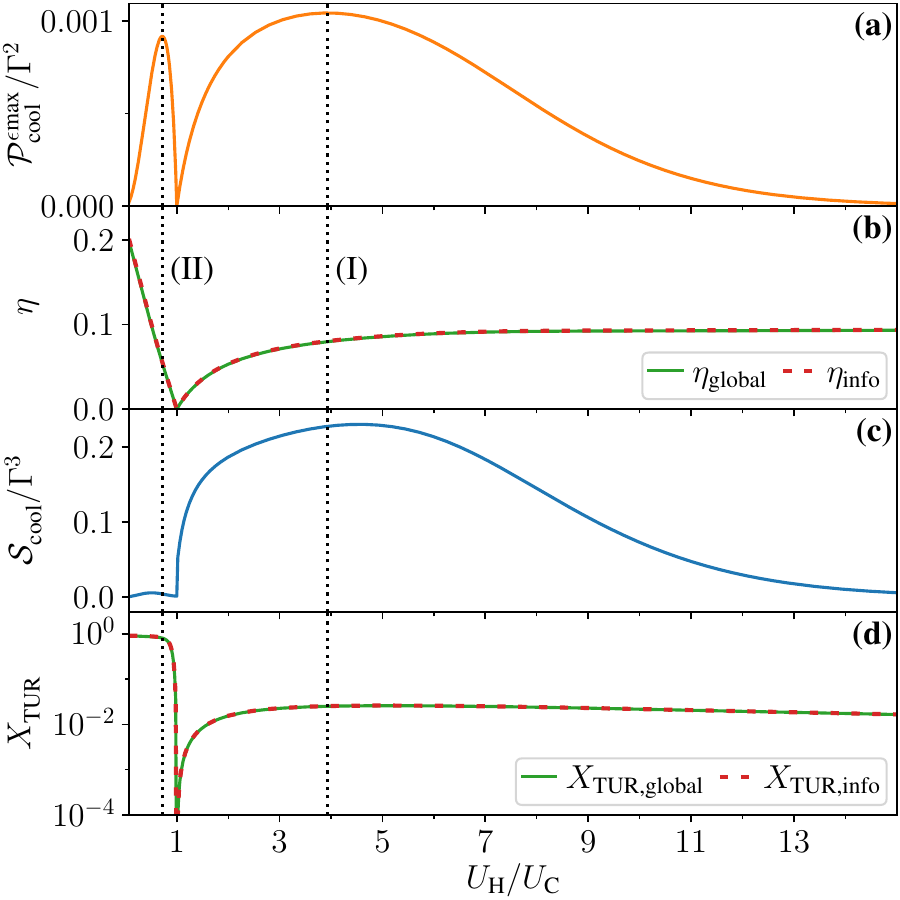}
    \caption{\label{fig:maxPcool_UH}
        (a) Cooling power maximized over $\eC$ and $\eM$, $\Pcool^{\epsilon\mathrm{max}}$, as a function of $\UH$ (see Appendix~\ref{app:power optimization} for the values of $\epsilon_\alpha$). We identify two local maxima (I) and (II), the position of which is indicated by dotted lines throughout the four panels of this figure.
        (b) Efficiencies $\eta_\mathrm{global}$ and $\eta_\mathrm{info}$, (c) fluctuations of the cooling power $\Scool$, and (d) precision trade-off parameters $X_\mathrm{TUR,global}$  and $X_\mathrm{TUR,info}$ at the same conditions as (a).
        In all panels, the level position of dot H, $\eH$ is set to fulfill the demon condition, $\Jin = 0$, and single-level energies $\eC$ and $\eM$ are fixed to maximize the cooling power. Other parameters (in units of $\GC =\GH =\Gamma$): $\THot = 16$, $\TCold = 4$, $\T = 8$, $\dT = 1$, $\UC=12$, and $\GL^{00} = \GR^{01} = \GR^{10} = 0.01$.
    }
\end{figure}

Typically, one of the most relevant performance quantifiers is the cooling power $\Pcool \equiv \JR$. Since the cooling power depends on all other parameters $\eC,\eM,\UC,\UH$, as well as $\TL,\TR,\THot,\TCold$, we choose to analyze the cooling power, $\Pcool$, maximized over all level positions $\eC,\eM$.
We show the result, the $\epsilon$-maximized cooling power, $\Pcool^{\epsilon\mathrm{max}}\equiv\mathrm{max}_{\epsilon_\H,\epsilon_\C}\Pcool$, in Fig.~\subfigref{fig:maxPcool_UH}{a} as a function of the ratio of interaction energies $\UH/\UC$ for a fixed interaction energy $\UC = 12$ and fixed temperatures $\THot = 16$, $\TCold = 4$, $\T = 8$, $\dT = 1$. All these values are expressed in units of the tunnel rate $\GC=\GH=\Gamma$ while $\GL^{00} = \GR^{01} = \GR^{10} = 0.01$. In other words, we here choose a situation in which the temperature difference in the resource part is larger than in the working substance.
Note that while one might imagine a trivial situation where exchange between the hot reservoirs of resource and working substance happens independently of heat exchange between the cold reservoirs of resource and working substance (which would hence also allow for a trivial interpretation of the demonic operation), this is not what we are considering here and it is even fully excluded by the fact that all energy exchange happens via the same dot W, which is indeed capacitively coupled to the full resource region. The resulting intricate energy exchange with the full resource region is furthermore reflected in the energy exchange of the basic cycles, as shown in Table~\ref{tab:cycles}, showing that all basic cycles exchange energy with three reservoirs, with a direction of energy flow that changes depending of the choice of parameters.

\begin{figure*}[ht!]
    \includegraphics[width=\linewidth]{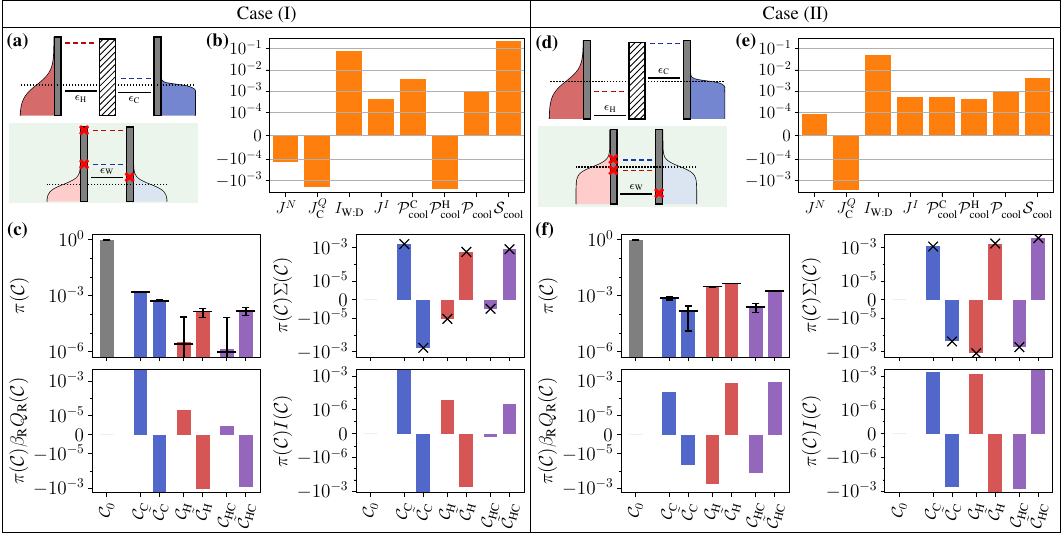}
    \caption{\label{fig:cycle occurrences}
        Analysis of the stochastic quantities for the basic cycles for two parameter configurations: Case (I) $\UH/\UC = 3.93$ and Case (II) $\UH/\UC = 0.72$, corresponding to the two local maxima of $\Pcool^{\epsilon\mathrm{max}}$, indicated by vertical dotted lines in Fig.~\subfigref{fig:maxPcool_UH}{a}. All other parameters are the same as in Fig.~\ref{fig:maxPcool_UH}.
        (a,d) Sketches of the level configurations of cases (I) and (II) (not at exact scale) and (b,e) bar plots characterizing the steady-state average currents, mutual information in the steady state $I_{\M:\D}$ and cooling power fluctuations $\Scool$ (divided by the adequate power of $\Gamma$ to make them dimensionless).
        (c,f) Stochastic thermodynamic quantities at the cycle level, see Table~\ref{tab:cycles}, weighted by the cycle probability $\pi(\cC)$ (plotted in the top left panel) to quantify their respective contribution to the average quantity.
        The probability $\pi(\cC)$ was obtained by counting the relative number of occurrences of cycle $\cC$ from (c) 65,000 / (f) 25,000 numerically generated trajectories of duration $2\times 10^5/\Gamma$.
        Note that a higher amount of trajectories was required for case (I) to match the analytical results since the occurrence of cycles involving the hot resource is strongly suppressed.
        The error bars in the top left panel correspond to the confidence interval $\pm 1/\sqrt{M_\text{cycles}}$, where $M_\text{cycles}$ is the total number of cycles found in all the trajectories combined, and the long black horizontal bars indicate the analytical value of $\pi(\cC)$ from Eq.~\eqref{eq:raterelation}. The crosses in the top right panel indicate the entropy production values computed from the numerical values of $\pi(\cC)$, namely $\log(\pi(\cC)/\pi(\bcC))$.
        Bar colors are chosen as a guide for the eye, matching the representation of the cycles in Fig.~\ref{fig:cycles}.
    }
\end{figure*}

The plot in Fig.~\subfigref{fig:maxPcool_UH}{a} shows two local maxima of the $\epsilon$-maximized cooling power, which are of a similar order of magnitude.
The larger of those two maxima is found for $\UH/\UC>1$, the slightly smaller one for $\UH/\UC<1$, indicating that the working principles of these two configurations, which we refer to as cases (I) and (II), are expected to be very different.
This expectation is confirmed by the trajectory analysis, which will be presented in Sec.~\ref{sec:discussion}, see also Fig.~\ref{fig:cycle occurrences}.

Interestingly, also the efficiencies, shown in Fig.~\subfigref{fig:maxPcool_UH}{b} are of similar magnitude at the positions (I) and (II), despite the fact that the underlying working principles are different. This applies both to the global efficiency $\eta_\mathrm{global}$, see Eq.~\eqref{eq:global_eff}, and to the information efficiency $\eta_\mathrm{info}$, see Eq.~\eqref{eq:info_eff}, which here turn out to be close to identical for any value of $\UH/\UC$. The reason for $\eta_\mathrm{global}\approx\eta_\mathrm{info}$ is that the Fourier heat flowing directly from hot to cold within the resource region is strongly suppressed in this parameter regime
\footnote{This is however not the case in other parameter regimes. In particular, for very large values of $\UC$ there can be a significant difference between the information and global performance quantifiers due to the entropy production in the resource region, see Appendix~\ref{app:power optimization}.}.
All in all, both cases are very similar in terms of average thermodynamic quantities, with even the heat currents from the cold bath $\JC$, the information flow $J^I$ and the mutual information in the steady-states $I_{\M:\D}$ being of the same order of magnitude, see Figs.~\subfigref{fig:cycle occurrences}{b} and \subfigref{fig:cycle occurrences}{e}.

In contrast to the similar \textit{average} performance of the two configurations of maximum cooling power, the difference in the precision is striking. This is shown in Fig.~\subfigref{fig:maxPcool_UH}{c}. Configuration (I), having a slightly larger maximum cooling power $\Pcool^{\epsilon\mathrm{max}}$, displays a maximum in cooling power fluctuations $\Scool$, which basically coincides with the maximum in $\Pcool^{\epsilon\mathrm{max}}$.
This means that while the desired output, namely the cooling power, in this configuration, is large, the precision is very low. This is in contrast with configuration (II), which has a slightly smaller maximum cooling power than (I), but at the same time strongly suppressed cooling power fluctuations $\Scool$.

This difference in performance becomes even more striking when considering the trade-off factor $X_\mathrm{TUR}$, shown in Fig.~\subfigref{fig:maxPcool_UH}{d}.
Indeed, this factor is not far from its bound under the demon condition , $X_\mathrm{TUR,global} \le 1$, for configuration (II), while for configuration (I) it is almost two orders of magnitude smaller.
It is furthermore interesting that the trade-off factor $X_\mathrm{TUR}$ exhibits two regions of almost plateau-like character, interrupted by the dip at $\UH=\UC$ at which the demon condition excludes a nonzero cooling power.
This clearly shows that the regime where the interaction energy $\UH$, mediating the energy exchange with the hot resource reservoir, is larger is far less advantageous from the point of view of the precision trade-off, with respect to the regime where the interaction energy $\UC$, mediating the energy exchange with the cold resource reservoir, is larger.

In the following, we analyze the different working principles of configuration (I) and (II), which are representative of the regimes $\UH>\UC$ and $\UC>\UH$, and show their impact on the performance with respect to average output and precision.
We note that having similar average performances in both configurations, in terms of cooling power and efficiency, is rather specific to our choice of interaction energy $\UC$. However, we emphasize that the two distinct working principles detailed in Sec.~\ref{sec:discussion} below apply to a wide range of values of $\UC$, as discussed in Appendix \ref{app:power optimization}.

\subsection{Distinguishing operation based on basic cycles}\label{sec:discussion}

The final important task is to identify in which way the working principles of the two optimal cases (I) and (II) differ, leading to very different performances with respect to the precision of the refrigeration process.

The addition energies of the three quantum dots in cases (I) and (II) result in the following situation, depicted in Figs.~\subfigref{fig:cycle occurrences}{a} and \subfigref{fig:cycle occurrences}{d}.
In case (I) the single-level energy of dot W is positive; a positive cooling power $\Pcool=\JR=-\JL$, see Eqs.~\eqref{JR2} and \eqref{JL2}, thus needs to go along with a negative particle current, $J^N<0$, cf. Fig.~\subfigref{fig:cycle occurrences}{b}, meaning that cooling is mediated by electron-like excitations.
By contrast, in case (II) the single-level energy of dot W is negative; a positive cooling power  $\Pcool>0$ thus needs to go along with $J^N>0$, cf. Fig.~\subfigref{fig:cycle occurrences}{e}, meaning that cooling is mediated by hole-like excitations
\footnote{When analyzing the two components of the heat current $\JR$ of Eq.~\eqref{JR2} instead, we find that the tightly coupled and the non-tightly coupled parts have opposite signs in both cases (I) and (II). For (I) both contributions have the same order of magnitude, with a positive tight-coupling contribution. For (II), the tight-coupling contribution is negative but much smaller than the non-tight-coupling one ($-\eM \sim \bar{U}$).}.

More important for the operation principle of the  device is the contribution of the hot resource reservoir to the cooling, quantified by $\Pcool^\mathrm{H}$, see Eq.~\eqref{Pcool^Ca} as well as Figs.~\subfigref{fig:cycle occurrences}{b} and \subfigref{fig:cycle occurrences}{e}. Indeed, in case (I), reservoir H acts detrimentally to the cooling, with $\Pcool^\mathrm{H} < 0$. Its main role in the device operation is therefore to provide the heat flow $\JH$ to suppress the average heat flow $\Jin$ from the demon to the working substance. Consequently, the cold resource reservoir is the only one to drive the cooling process, in the same manner as described in Ref.~\cite{Strasberg2013Jan} for a double-dot setup with a single-reservoir resource. Conversely, in case (II), both resource reservoirs contribute positively to the cooling process, \textit{collaborating} hence in parallel. While this consideration concerns an effect of the cooling power contribution from each resource contact (namely the decomposition in terms of cycles of the fundamental set, as given in Eq.~\eqref{current split}), also the contribution of the different basic cycles, as introduced in Sec.~\ref{sec:basic cycles}, to the operation plays an important role as discussed in the following.

Indeed, an interesting aspect distinguishing cases (I) and (II) is the information exchange as compared to the heat exchange going along with the different cycles.
The basis to understand the importance of the acquired information is the positions of the addition energies of the resource dots, shown in Figs.~\subfigref{fig:cycle occurrences}{a} and \subfigref{fig:cycle occurrences}{d}. In case (I), the Coulomb interaction of dot H is large. As a result of this and of the position of the single-level energies, both dots H and C hence ``see'' both the electron-like and the hole-like parts of the distributions of the reservoirs they are tunnel-coupled to.
The two addition energies are thus aligned with very different parts of the distributions, making both of them effectively cold (note that $U_\H\approx3T_\H$).
This is in contrast to case (II), where the addition energies of dot H communicate only with (mostly empty) hole-like excitations of reservoir H and dot C only communicates with electron-like excitations of reservoir C. Whether dot W is occupied or not hence has a smaller influence on the tunnel rates of dot H.
This is visible particularly clearly in the information exchanged with the basic cycle $\CH$: $I(\CH) \ll I(\CC)$ in case (II), while $I(\CH) \simeq I(\CC)$ in case (I); see Fig.~\ref{fig:cycle occurrences} and also Fig.~\ref{fig:cycle unweighted} in Appendix \ref{app:cycle unweighted} to compare the information (as well as other thermodynamic quantities) at the cycle level.
Furthermore, in case (I) both $\CC$ and $\CH$, which have the largest positive cooling power of the three basic cycles, go along with information received by the demon, see Fig.~\subfigref{fig:cycle occurrences}{c}. Interestingly in case (II), while cycles $\CC$ and $\bCHC$ exploit an information flow into the demon, this is not the case for cycle $\CH$, whose contribution to the cooling power is negative, see Fig.~\subfigref{fig:cycle occurrences}{f}. The beneficial working principle of cycle $\CH$ is hence more similar to the one of an absorption refrigerator, using heat for a useful task, such as in the double-dot heat engines of Refs.~\cite{Sanchez2011Feb,Thierschmann2015Oct}, see also a qubit implementation~\cite{Bhandari2021Aug}.

A further important difference between the two cases, which has an influence on the fluctuations of the cooling power is the following.
When analyzing the probability with which basic cycles $\cC$ occur compared to their time-reversed $\bcC$, we note that in case (I), the more probable $\bCH$ and  $\bCHC$ contribute with a negative cooling power. They are hence counterproductive to the result of cycle $\CC$, which is more probable than $\bCC$ and contributes with a large positive cooling power.
The situation is very different in case (II). Here the more probable of the pairs of basic cycles $\cC$ and $\bcC$ ($\CC$, $\CH$, and $\tilde\CHC$ in this case), are all beneficial for the refrigeration process. It is clear that this reduces significantly the fluctuations in the cooling power.

\section{Conclusions}
\label{sec:conclusions}

We have provided a detailed analysis of an information-based refrigerator exploiting capacitive coupling between three quantum dots, inspired by previous proposals and realizations of quantum-dot-based Maxwell demons~\cite{Strasberg2013Jan,Koski2015Dec} and heat engines~\cite{Sanchez2011Feb,Thierschmann2015Oct}.

We showed that this triple-dot  refrigerator can work as an ideal autonomous Maxwell demon generating finite cooling power in the working substance, while not extracting any heat on average from the resource region.
We showed furthermore that the precision of this refrigerator and the trade-off between precision, cooling power, and efficiency, strongly depends on the underlying parameter regimes, in particular on the ratio of interaction energies with different parts of the resource region.

Our detailed analysis based on steady-state master equation dynamics, full counting statistics, as well as trajectory-resolved dynamics allowed us to identify that refrigerator realizations mostly exploiting information suffer from reduced precision compared to those realizations which exploit both information and heat engine aspects.

We expect these insights to be important for the design of energy-converting devices, which exploit generic types of resources that are not characterized by one unique temperature.\\

The data that support the findings of this article are openly available \cite{Data}.

\acknowledgments
We are grateful for very helpful and inspiring discussions with Robert Whitney. We thank Abhaya Hegde for useful comments on the manuscript. We acknowledge financial support from the Knut and Alice Wallenberg foundation through the fellowship program (M.A.,J.M.,J.S.), from the European Research Council (ERC) under the European Union’s Horizon Europe research and innovation program (101088169/NanoRecycle) (J.S.) and from the Spanish Ministerio de Ciencia e Innovaci\'on via grants No. PID2019-110125GB-I00 and No. PID2022-142911NB-I00, and through the ``Mar\'{i}a de Maeztu'' Programme for Units of Excellence in R{\&}D CEX2023-001316-M (R.S.).

\appendix

\section{Steady state} \label{app:ss}

\subsection{Analytical solution}
For the case considered in Sec.~\ref{sec:case},  $\GR^{00} = \GL^{10} = \GL^{01} = 0$, we solve the equation $W\bar{\P} = 0$ and obtain the steady state
\allowdisplaybreaks
\begin{widetext}
\small
\begin{equation*}\begin{split}
    \p_{000} = \frac{1}{\g} \Big(
          \GCom\GCim\GHom\GHim\GLoom + \GCom\GCip\GHom\GHim\GRoim + \GCom\GHom\GHim\GLoom\GRoim + \GCim\GHom\GHim\GLoom\GRoip + \GCom\GCim\GHom\GHip\GRiom \\
        + \GCom\GCim\GHom\GLoom\GRiom + \GCom\GCip\GHom\GRoim\GRiom + \GCom\GHom\GHip\GRoim\GRiom + \GCom\GHom\GLoom\GRoim\GRiom + \GCim\GHom\GHip\GRoip\GRiom   \\
        + \GCim\GHom\GLoom\GRoip\GRiom + \GCom\GCim\GHim\GLoom\GRiop + \GCom\GCip\GHim\GRoim\GRiop + \GCom\GHim\GLoom\GRoim\GRiop + \GCim\GHim\GLoom\GRoip\GRiop\Big)
\end{split}\end{equation*}
\begin{equation*}\begin{split}
    \p_{001} = \frac{1}{\g} \Big(
          \GCop\GCim\GHom\GHim\GLoom + \GCop\GCip\GHom\GHim\GRoim + \GCop\GHom\GHim\GLoom\GRoim + \GCip\GHom\GHim\GLoop\GRoim + \GCop\GCim\GHom\GHip\GRiom \\
        + \GCop\GCim\GHom\GLoom\GRiom + \GCop\GCip\GHom\GRoim\GRiom + \GCop\GHom\GHip\GRoim\GRiom  + \GCop\GHom\GLoom\GRoim\GRiom + \GCip\GHom\GLoop\GRoim\GRiom \\
        + \GCop\GCim\GHim\GLoom\GRiop + \GCop\GCip\GHim\GRoim\GRiop + \GCip\GHop\GHim\GRoim\GRiop + \GCop\GHim\GLoom\GRoim\GRiop + \GCip\GHim\GLoop\GRoim\GRiop\Big)
\end{split}\end{equation*}
\begin{equation*}\begin{split}
    \p_{010}  = \frac{1}{\g} \Big(
          \GCom\GCim\GHom\GHim\GLoop + \GCom\GHom\GHim\GLoop\GRoim + \GCop\GCim\GHom\GHim\GRoip + \GCim\GHom\GHim\GLoop\GRoip + \GCom\GCim\GHom\GLoop\GRiom \\
        + \GCom\GHom\GLoop\GRoim\GRiom + \GCop\GCim\GHom\GRoip\GRiom + \GCim\GHom\GLoop\GRoip\GRiom  + \GCom\GCim\GHop\GHim\GRiop + \GCom\GCim\GHim\GLoop\GRiop \\
        + \GCom\GHop\GHim\GRoim\GRiop + \GCom\GHim\GLoop\GRoim\GRiop + \GCop\GCim\GHim\GRoip\GRiop + \GCim\GHop\GHim\GRoip\GRiop + \GCim\GHim\GLoop\GRoip\GRiop\Big)
\end{split}\end{equation*}
\begin{equation*}\begin{split}
    \p_{011}  = \frac{1}{\g} \Big(
          \GCom\GCip\GHom\GHim\GLoop + \GCop\GCip\GHom\GHim\GRoip + \GCop\GHom\GHim\GLoom\GRoip + \GCip\GHom\GHim\GLoop\GRoip  + \GCom\GCip\GHom\GLoop\GRiom \\
        + \GCop\GCip\GHom\GRoip\GRiom + \GCop\GHom\GHip\GRoip\GRiom + \GCop\GHom\GLoom\GRoip\GRiom  + \GCip\GHom\GLoop\GRoip\GRiom + \GCom\GCip\GHop\GHim\GRiop \\
        + \GCom\GCip\GHim\GLoop\GRiop + \GCop\GCip\GHim\GRoip\GRiop  + \GCip\GHop\GHim\GRoip\GRiop + \GCop\GHim\GLoom\GRoip\GRiop + \GCip\GHim\GLoop\GRoip\GRiop\Big)
\end{split}\end{equation*}
\begin{equation*}\begin{split}
   \p_{100}  = \frac{1}{\g} \Big(
           \GCom\GCim\GHop\GHim\GLoom + \GCom\GCip\GHop\GHim\GRoim + \GCom\GHop\GHim\GLoom\GRoim + \GCim\GHop\GHim\GLoom\GRoip + \GCom\GCim\GHop\GHip\GRiom \\
         + \GCom\GCim\GHop\GLoom\GRiom + \GCom\GCim\GHip\GLoop\GRiom + \GCom\GCip\GHop\GRoim\GRiom   + \GCom\GHop\GHip\GRoim\GRiom + \GCom\GHop\GLoom\GRoim\GRiom \\
         + \GCom\GHip\GLoop\GRoim\GRiom + \GCop\GCim\GHip\GRoip\GRiom   + \GCim\GHop\GHip\GRoip\GRiom + \GCim\GHop\GLoom\GRoip\GRiom + \GCim\GHip\GLoop\GRoip\GRiom\Big)
\end{split}\end{equation*}
\begin{equation}\begin{split}
   \p_{110}  = \frac{1}{\g} \Big(
          \GCom\GCim\GHom\GHip\GLoop + \GCom\GHom\GHip\GLoop\GRoim + \GCop\GCim\GHom\GHip\GRoip + \GCim\GHom\GHip\GLoop\GRoip+ \GCom\GCim\GHop\GHip\GRiop \\
        + \GCom\GCim\GHop\GLoom\GRiop + \GCom\GCim\GHip\GLoop\GRiop + \GCom\GCip\GHop\GRoim\GRiop  + \GCom\GHop\GHip\GRoim\GRiop + \GCom\GHop\GLoom\GRoim\GRiop \\
        + \GCom\GHip\GLoop\GRoim\GRiop + \GCop\GCim\GHip\GRoip\GRiop   + \GCim\GHop\GHip\GRoip\GRiop + \GCim\GHop\GLoom\GRoip\GRiop + \GCim\GHip\GLoop\GRoip\GRiop\Big)
\end{split}\label{steady state}
\end{equation}
\end{widetext}
where $\g$ is the normalization factor required to have $\sum_{hwc}\p_{hwc}=1$. This result for the steady-state probabilities allows us to get the expressions of the average currents from Eq.~\eqref{average_current}. For each kind of current, $\nu = N, E, Q, S, I$, we then check that the obtained expression is equal to the basic-cycle-based expression given by Eq.~\eqref{current}.

The steady-state occupations, $\bar{\P}$, and currents can also be computed graphically by applying network theory results \cite{Schnakenberg1976Oct} to the Markov chain graph representing the transitions in the studied regime, as explained in the next section.

\subsection{Network theory analysis of the master equation}\label{app:graphs}

\begin{figure*}
    \includegraphics[width=\linewidth]{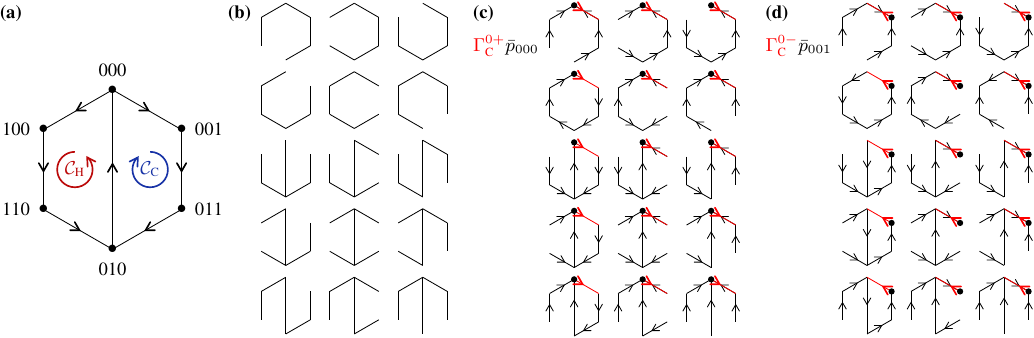}
    \caption{\label{fig:graphs}
        Illustration of the graphical calculation of the steady state probabilities and fluxes.  (a) Graph representation $G$ (with an arbitrary choice of direction for each edge, see Ref.~\cite{Schnakenberg1976Oct}) of the master equation \eqref{eq:master} and chosen set of fundamental cycles, $\{\CC, \CH\}$. (b) Set of maximal trees of $G$, $\{T^{(i)}(G)\}_{i=1,..,15}$. (c), (d) Graphical calculation of the flux $\FC$: sets of maximal trees of $G$ (in black) directed respectively towards states $000$ and $001$ used in the calculation of $\p_{000}$ and $\p_{001}$. The added red edge in each graph corresponds to the extra rate, respectively $\GCop$ and $\GCom$, appearing in the first and second terms of $\FC$, cf. Eq.~\eqref{F(CC)}.
    }
\end{figure*}

Applying the methods from Ref.~\cite{Schnakenberg1976Oct} to the triple-dot setup in the case considered in Sec.~\ref{sec:case}, we obtain the graph $G$ depicted in Fig.~\subfigref{fig:graphs}{a}, which is similar to the Markov chain representation from Fig.~\subfigref{fig:setup}{b} but replacing each pair of arrows representing the transitions by a single edge with an arbitrary direction, corresponding to our choice of direction for cycles  $\CC$, $\CH$, see Fig.~\ref{fig:cycles}. A fundamental set of cycles for $G$ is $\{\CC, \CH\}$. There are other possible sets but this choice is physically more meaningful than, e.g., $\{\CC, \CHC\}$ since it allows us to separate the contribution from the hot and cold resource reservoirs to the thermodynamic quantities.

We can then construct the steady state $\bP$ from the set of maximal trees of $G$, $\{T^{(i)}(G)\}_{i=1,..,15}$ given in Fig.~\subfigref{fig:graphs}{b}. A maximal tree $T(G)$ is a subgraph of $G$ consisting of all the vertices of $G$ and a subset of edges of $G$, such that $T(G)$ is connected and contains no cycles.
The steady state is then given by $\p_{hwc} = \gamma_{hwc}^5/\g$ \cite{Schnakenberg1976Oct}, where
\begin{equation}
     \gamma_{hwc}^5 = \sum_{i=1}^{15} g\bm{(}T_{hwc}^{(i)}(G)\bm{)},\quad\g = \sum_{h,w,c}\gamma_{hwc}^5.
     \label{eq:tree_ss}
\end{equation}
$T_{hwc}^{(i)}(G)$ is the $i$-th maximal tree of $G$ in which all the edges are directed toward state $hwc$, see Figs.~\subfigref{fig:graphs}{c} and \subfigref{fig:graphs}{d} for the sets of directed maximal trees for states $000$ and 001 respectively. Then, the quantity $g\bm{(}T_{hwc}^{(i)}(G)\bm{)}$ associated to the tree $T_{hwc}^{(i)}(G)$ is the product of the transition rates $W_{h'w'c'}^{h''w''c''}$ corresponding to the directed edges $h''w''c''\to h'w'c' $ of $T_{hwc}^{(i)}(G)$, see also Fig.~\subfigref{fig:setup}{b}. From Eq.~\eqref{eq:tree_ss}, we find the same results as in Eq.~\eqref{steady state}.

Furthermore, the entropy production rate in the steady state can be expressed in terms of affinities $A$ and fluxes $F$ for each cycle of the chosen fundamental set \cite{Schnakenberg1976Oct, Horowitz2014Jul},
\begin{equation}
    \dot{\Sigma} = A^\C\FC + A^\H\FH,
\end{equation}
where, in our case,
\begin{subequations}
\begin{align}
     A^\alpha &= \log\left(\frac{\gamma_{\cC_\alpha}^4}{\gamma_{\bcC_\alpha}^4}\right), \\
     \FC &= \GCop\p_{000} - \GCom\p_{001}, \label{F(CC)}\\
     \FH &= \GHop\p_{000} - \GHom\p_{100},\label{F(CH)}
\end{align}
\end{subequations}
with $\alpha = \C,\H$.  Using the analytical expression \eqref{steady state}, one can show that the expressions \eqref{F(CC)} and \eqref{F(CH)} are equal to the fluxes $\FC$ and $\FH$ given in Eq.~\eqref{F(C)} in the main text. Alternatively, these fluxes can be computed graphically, as illustrated in Fig.~\ref{fig:graphs} for $\FC$. Figures~\subfigref{fig:graphs}{c} and \subfigref{fig:graphs}{d} respectively show the trees corresponding to the first and second term of Eq.~\eqref{F(CC)}, the edge in red corresponding to the added transition rate in front of the probability $\p_{hwc}$, namely $000 \to 001$ for $\GCop$ and  $001 \to 000$ for $\GCom$, see also Fig.~\subfigref{fig:setup}{b}. We see that most of the trees (including the red edge) cancel each other in Eq.~\eqref{F(CC)}, except for the four trees where the red edge was not already an edge of the tree (disregarding the orientation), leading to Eq.~\eqref{F(C)}. Then, from the definition of the stochastic entropy production \eqref{eq:FR}, we obtain $A^\alpha = \Sigma(\cC_\alpha)$, with $\alpha = \C,\H$, leading to
the entropy production rate
\begin{equation}\label{dSig1}
     \dot{\Sigma} = \Sigma(\CC)\FC + \Sigma(\CH)\FH.
\end{equation}
Finally, using Eq.~\eqref{F(C)}, $\Sigma(\bcC) = -\Sigma(\cC)$ and $\Sigma(\CHC) = \Sigma(\CH) - \Sigma(\CC)$, we get
\begin{equation}\label{dSig2}
    \dot{\Sigma} = \sum_{\cC=\CC, \CH, \cC_{\H\C}} [\Sigma(\cC) r_{\cC} +\Sigma(\bcC) r_{\bcC}].
\end{equation}

With the same kind of graphical analysis and using the expression of the current, Eq.~\eqref{average_current}, we find the equivalent of Eqs.~\eqref{dSig1} and \eqref{dSig2} for the steady-state current $J^\nu_\alpha$, namely Eqs.~\eqref{current split} and \eqref{current}.

\section{Thermodynamic uncertainty relation} \label{app:TUR}

Since we are studying the steady state of a Markov jump process, the thermodynamic uncertainty relation holds \cite{Gingrich2016Mar}, namely
\begin{equation}\label{TUR}
   \Scool\dot{\Sigma} \ge 2 \Pcool^2.
\end{equation}
Based on this relation, we define the coefficient
\begin{equation}
    x_\text{TUR, global} = \frac{2 \Pcool^2}{\Scool\dot{\Sigma}},
\end{equation}
which is therefore bounded by 1. Using the expression $\etaG$, Eq.~\eqref{eq:global_eff}, we get
\begin{equation}
    \frac{\etaG}{1- \etaG} = \frac{\bL\JL + \bR\JR}{\dot{\Sigma}},
\end{equation}
such that
\begin{equation}
     x_\text{TUR, global} = \frac{2 \Pcool}{\Scool}\frac{\etaG}{1- \etaG} \frac{\JR}{\bL\JL + \bR\JR},
\end{equation}
since $\Pcool = \JR$. We are interested here in the case where there is no potential bias, $\mu_\R = 0$, and the demon condition is fulfilled, therefore $\JL = -\JR$. With this, we find
\begin{equation}
    x_\text{TUR, global} = \frac{2 \Pcool}{\Scool}\frac{\etaG}{1- \etaG} \frac{\TL\TR}{\TL-\TR},
\end{equation}
which is our motivation for the definition of the performance quantifier $X_\text{TUR, global}$ given in Eq.~\eqref{X_TUR}.

\begin{figure}[b]
    \includegraphics[width=\linewidth]{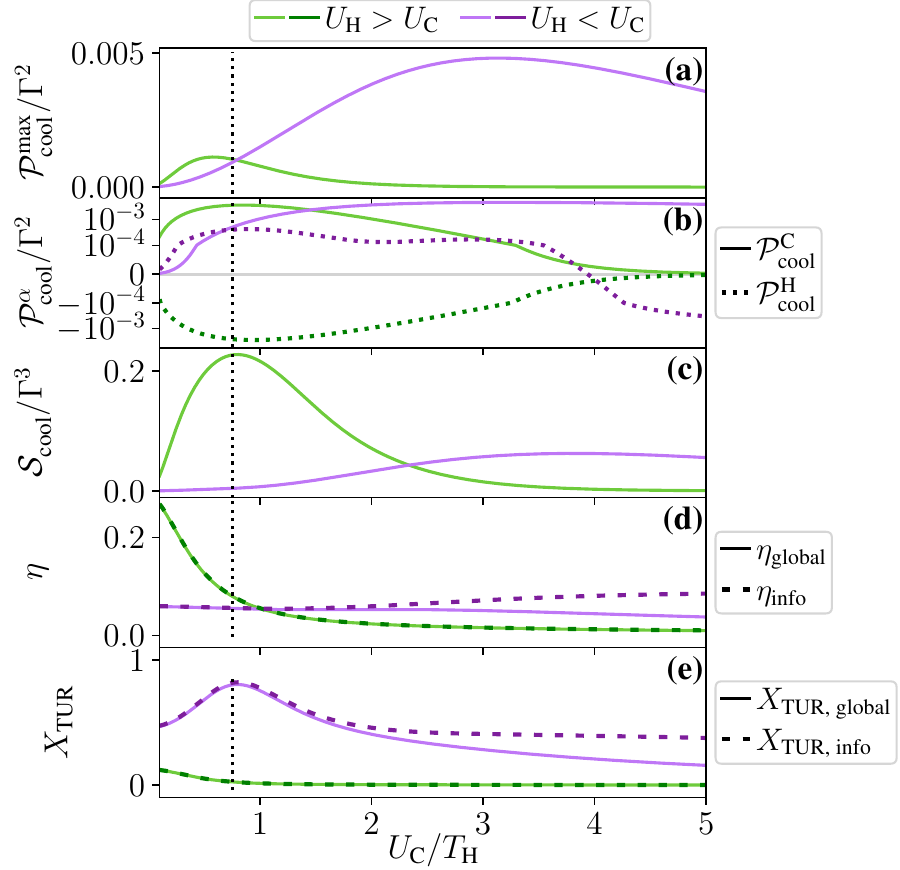}
    \caption{\label{fig:app:maxPcool_UC}
        (a) $\Pcool^\text{max}$, the cooling power maximized over $\eC$, $\eM$ and $\UH$, with the constraint $\UH > \UC$ (green) or $\UH < \UC$ (purple curves).
        (b) Components $\Pcool^{\C}$ and $\Pcool^{\H}$ of $\Pcool^{\mathrm{max}}$.
        (c) Cooling power fluctuations $\Scool$.
        (d) Efficiencies $\eta_\mathrm{global/info}$.
        (e) Precision trade-off parameters $X_\mathrm{TUR,global/info}$.
        All the quantities are plotted as functions of $\UC$, at the values of $\eC$, $\eM$, and $\UH$ maximizing the cooling power.
        The vertical dotted black line indicates the value of $\UC$ used in all the other figures. The other parameters are the same as in Fig.~\ref{fig:maxPcool_UH}.
    }
\end{figure}

By analogy, we define
$x_\text{TUR, info} = {2 \Pcool^2}/{\Scool\dot{\Sigma}_\M}$,
which becomes $X_\text{TUR, info}$ at $\mu_\R = 0$, $\Jin = 0$. However, the reduced dynamics of dot W alone cannot be described by a Markov jump process due to the capacitive couplings to dots C and H. Therefore, the TUR does generally not hold and $x_\text{TUR, info}$ is not bounded by 1. Nevertheless, we observe in our plots (Figs.~\subfigref{fig:maxPcool_UH}{d} and \subfigref{fig:app:maxPcool_UC}{e}) that  $X_\text{TUR, info}\le 1$ for our choice of parameters, such that there is no TUR violation for the reduced system.

Also note that at the demon condition and $\mu_\R = 0$, the global efficiency from Eq.~\eqref{eq:global_eff} can be expressed as
\begin{equation}
    \etaG = \frac{\bR - \bL}{\bC - \bH}\frac{\eM(\UH - \UC)}{\UC\UH} \le 1,
\end{equation}
showing that the efficiency increases with the temperature difference $\dT$ between the working substance reservoirs but decreases when increasing the temperature difference between the resource reservoirs. However, note that there is a maximum value of $\dT$ above which the demon condition can no longer be achieved.

\begin{figure}[b]
    \includegraphics[width=\linewidth]{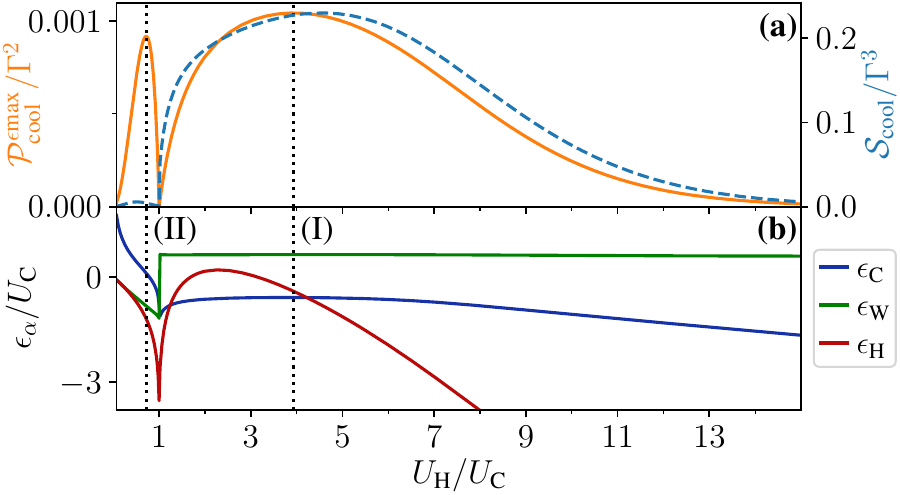}
    \caption{\label{fig:app:maxPcool_UH}
        (a) Plot of $\Pcool^{\epsilon\mathrm{max}}$ (solid orange line), the cooling power $\Pcool$ maximized over $\eC$ and $\eM$, as a function of $\UH$ and corresponding fluctuations of the cooling power $\Scool$ (dashed blue line).
        (b) Single-level energies $\eC$ and $\eM$ maximizing the cooling power and energy $\eH$ giving the demon condition, $\Jin = 0$.
        The parameters are the same as in Fig.~\ref{fig:maxPcool_UH}.
    }
\end{figure}

\section{Mutual information and entropy flow at the trajectory  level} \label{app:extra info qties trajectory}

The stochastic mutual information along trajectory $\gamma$ can be defined as follows:
\begin{equation}
    \cI_{\M:\D}(\gamma) = \log\left(\frac{\p(\gamma_M)}{\p^\M_{w_M}\p^\D_{h_M  c_M}}\right) -  \log\left(\frac{\p(\gamma_0)}{\p^\M_{w_0}\p^\D_{h_0  c_0}}\right).
\end{equation}
Note that we here have $\cI_{\M:\D}(\gamma) = \cI_\M(\gamma) + \cI_\D(\gamma)$, with $\cI_\M(\gamma)$ and $\cI_\D(\gamma)$ given in Eq.~\eqref{info traj}. This is proven using $p(x|y) = p(x,y)/p(y)$ combined with the sequential tunneling regime.

Since transitions are only due to sequential tunneling, we can further split $\cI_\D(\gamma) = -S_\C(\gamma) -S_\H(\gamma)$ with
\begin{gather}
\begin{aligned}
    S_\C(\gamma) &= -\log\left(\prod_{m=1}^M\frac{\p(w_{m-1}|h_{m-1},c_{m})}{\p(w_{m-1}| h_{m-1},c_{m-1})}\right),\\
    S_\H(\gamma) &= -\log\left(\prod_{m=1}^M\frac{\p(w_{m-1}|h_{m},c_{m-1})}{\p(w_{m-1}| h_{m-1},c_{m-1})}\right),
\end{aligned}
\end{gather}
and  similarly, $\cI_\M(\gamma) = -S_\L(\gamma)  -S_\R(\gamma)$, by counting only the jumps mediated by the left---respectively the right---reservoir in Eq.~\eqref{info traj}.
$S_\alpha(\gamma)$ corresponds to the entropy current $J^S_\alpha$ integrated over the trajectory $\gamma$.

\section{Power optimization} \label{app:power optimization}

\begin{figure*}[tb]
    \includegraphics[width=\linewidth]{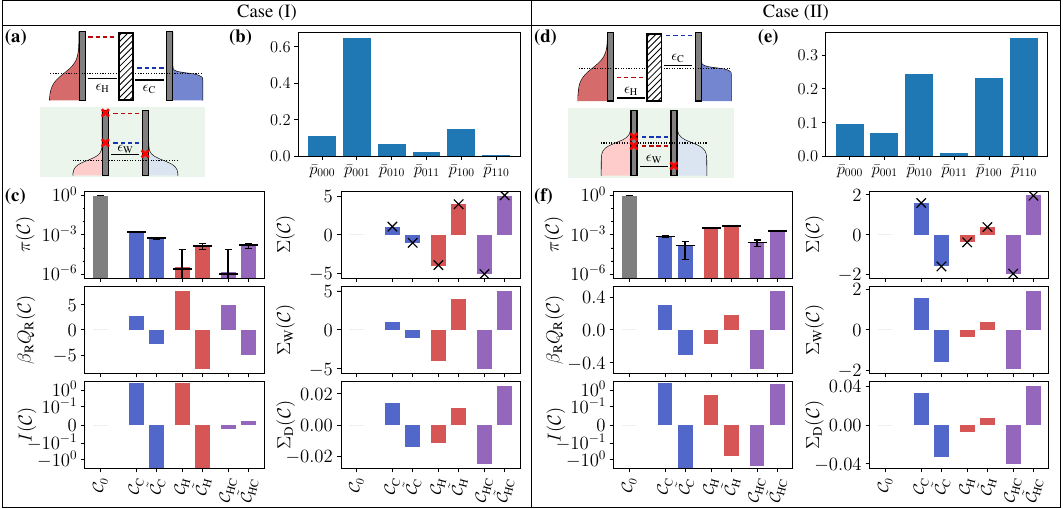}
    \caption{\label{fig:cycle unweighted}
        Analysis of the stochastic quantities for the basic cycles for cases (I) and (II) supplementing Fig.~\ref{fig:cycle occurrences}.
        (a,d) Sketches of level configurations of cases (I) and (II) (not at exact scale) and (b,e) bar plots of the steady-state probabilities $\p_{hwc}$.
        (c,f) Stochastic thermodynamic quantities at the cycle level, see Table~\ref{tab:cycles} and Eqs.~\eqref{stoc loc entropy prod}. Unlike in Fig.~\ref{fig:cycle occurrences}, these quantities are not weighted by the cycle probability $\pi(\cC)$, which is plotted in the top left panel.
        The long black horizontal bars in that panel indicate the analytical value of $\pi(\cC)$ from Eq.~\eqref{eq:raterelation}. The crosses in the top right panel indicate the entropy production values computed from the numerical values of $\pi(\cC)$, namely $\log(\pi(\cC)/\pi(\bcC))$.
    }
\end{figure*}

To supplement Fig.~\ref{fig:maxPcool_UH}, together with the optimized cooling power and its noise in Fig.~\subfigref{fig:app:maxPcool_UH}{a} for reference, we have plotted the single-level energies $\eC$ and $\eM$ maximizing the cooling power and the energy $\eH$ giving the demon condition, $\Jin = 0$, as functions of $\UC/\UH$ in Fig.~\subfigref{fig:app:maxPcool_UH}{b}. We see also clearly in this plot the two different operating regimes, $\UH > \UC$, like in case (I), and $\UH < \UC$, like in case (II), with a sharp change at $\UC = \UH$. For $\UH > \UC$, the optimal values of $\eC$, $\eM$ are only weakly dependent on $\UH$, like the efficiencies [Fig.~\subfigref{fig:maxPcool_UH}{b}] and precision trade-off parameters [Fig.~\subfigref{fig:maxPcool_UH}{d}]. On the contrary, for $\UH < \UC$, $\eC$, $\eM$ and the efficiencies $\eta_\mathrm{global/info}$ steeply decrease when $\UH$ gets closer to $\UC$ while $X_\mathrm{TUR,global/info}$ remain constant almost until $\UH$ reaches $\UC$.

In the main text, we have discussed the operating mode and performances of the device based on the choice of the interaction energy $\UH$, cases (I) and (II). For our choice of other parameters (temperatures, tunnel couplings, and interaction energy $\UC$) we found that cases (I) and (II) have similar cooling powers and efficiencies but very different precisions due to the difference in operating mode (see Sec.~\ref{sec:discussion}). We investigate here how dependent on the value of $\UC$ those findings are by plotting in Fig.~\ref{fig:app:maxPcool_UC} the key performance quantifiers of the refrigerator as functions of $\UC$ for $\UH > \UC$ (green curves) and $\UH < \UC$ (purple curves). More precisely, we are interested in the maximized cooling powers
\begin{gather}
\begin{aligned}
    \Pcool^{\text{max}, \UH>\UC} &= \max_{\eC,\eM, \UH\ge\UC}\Pcool,\\
    \Pcool^{\text{max}, \UH<\UC} &= \max_{\eC,\eM, 0\le \UH\le\UC}\Pcool,
\end{aligned}
\end{gather}
respectively corresponding to cases (I) and (II) when $\UC = 12\Gamma$ (vertical dotted black line in Fig.~\ref{fig:app:maxPcool_UC}). In Figs.~\subfigref{fig:app:maxPcool_UC}{a} and \subfigref{fig:app:maxPcool_UC}{d}, we see that having $\eta_\text{global/info}^{\UH>\UC} \simeq \eta_\text{global/info}^{\UH<\UC}$ and $\Pcool^{\text{max,}\UH>\UC} \simeq \Pcool^{\text{max,}\UH<\UC}$ is restricted to a small range of values of $\UC$, which also coincide with the maximal values of $X_\text{TUR,global/info}$ [Fig.~\subfigref{fig:app:maxPcool_UC}{e}]. However, the identified operating modes apply to a much larger range of values. This is shown by splitting $\Pcool^{\text{max,}\UH\gtrless\UC}$ into the two components $\Pcool^{\C,\UH\gtrless\UC}$ and $\Pcool^{\H,\UH\gtrless\UC}$ from the cold and hot resource reservoirs. Figure~\subfigref{fig:app:maxPcool_UC}{b} shows that  $\Pcool^{\C, \UH>\UC}$ is always negative, that is the hot resource reservoir works against the cooling, while $\Pcool^{\H, \UH<\UC}$ and $\Pcool^{\C, \UH<\UC}$ are both positive for $\UC$ up to around $4\THot$ and therefore both resource reservoirs contribute in parallel to the cooling. Overall, the performances of the device, combining cooling power, efficiency and precision, as quantified by $X_\text{TUR,global/info}$, are always significantly larger in the case $\UH < \UC$. Note that $X_\text{TUR,global}$ remains farther away from its bound, 1, than in Fig.~\subfigref{fig:maxPcool_UH}{d} since the maximum of $X_\text{TUR,global}$ is not reached for the same value of $\UH$ as the maximum cooling power.
Finally, in the case $\UH < \UC$, the efficiencies $\etaG$ and $\etaI$ become increasingly different at larger $\UC$ [Fig.~\subfigref{fig:app:maxPcool_UC}{d}] due to an increased entropy production in the resource region. This makes in turn $X_\text{TUR,info}$ deviate from $X_\text{TUR,global}$ [Fig.~\subfigref{fig:app:maxPcool_UC}{e}].\\

\section{Thermodynamic quantities at the cycle level} \label{app:cycle unweighted}

We characterize more in-depth the differences between cases (I) and (II), depicted in  Figs.~\subfigref{fig:cycle unweighted}{a} and \subfigref{fig:cycle unweighted}{d}, by plotting the corresponding steady-state probabilities in Figs.~\subfigref{fig:cycle unweighted}{b} and \subfigref{fig:cycle unweighted}{e}. We see that in case (I), the system is most often in state 001 while case (II) has a more uniform probability distribution, with most $\p_{hwc}$ of the order of 0.1-0.2. Figs.~\subfigref{fig:cycle unweighted}{c} and \subfigref{fig:cycle unweighted}{f} show the same stochastic thermodynamic quantities at the cycle level as in Fig.~\ref{fig:cycle occurrences}, with the addition of the local entropy productions $\Sigma_{\M}$ and $\Sigma_\D$, but without weighting them with the corresponding cycle probability $\pi(\cC)$. We can see that during cycles $\CC$ and $\CH$, the demon acquires an almost identical amount of information on dot W in case (I), which is consistent with the fact that even the hot demon reservoir appears as ``cold'' in this regime.
Conversely, in case (II), much less information is acquired during $\CH$ and it is actually $\bCH$ which contributes to the cooling ($Q_\R(\bCH) > 0$) while the demon provides information to dot W ($I(\bCH) < 0$), highlighting that the cooling is done thanks to the thermal resource of the hot reservoir in that case. Finally, as expected from Fig.~\subfigref{fig:maxPcool_UH}{b} which shows that $\etaG \simeq \etaI$, in both cases $\Sigma(\cC) \simeq \Sigma_\M(\cC)$, namely most of the entropy production is generated in the working substance. Note that the local entropy production $\Sigma_\M$ contains not only the entropy exchanged with the left and right reservoirs but also the information exchanged with the demon, see Eq.~\eqref{stoc loc entropy prod}.

\bibliography{main.bib}

\end{document}